\documentclass{aa}  

\usepackage{graphicx}
\usepackage{comment}
\usepackage{xcolor}
\usepackage{multirow}
\usepackage{txfonts}
\usepackage[colorlinks=true,linkcolor=blue,citecolor=blue]{hyperref}%
\newcommand{\OIII}{{[O\,{\sc iii}]\,}}

\newcommand{\OIIIl}{{[O\,{\sc iii}]\,$\lambda$}}

\newcommand{\Ha}{H$\alpha$\,}

\newcommand{\Hb}{H$\beta$\,}

\begin{document}

   \title{The radio properties of the JWST-discovered AGN}

   \subtitle{}

   \author{G. Mazzolari
          \inst{1,2}\fnmsep
          \thanks{giovanni.mazzolari@inaf.it}
          \and
          R. Gilli\inst{2}
          \and 
          R. Maiolino\inst{3,4,5}
          \and 
          I. Prandoni\inst{6}
          \and
          I. Delvecchio\inst{2}
          \and
          C. Norman\inst{7}
          \and
          E.~F. Jim\'enez-Andrade \inst{10}
          \and
          S. Belladitta \inst{2,11}
          \and
          F. Vito\inst{2}
          \and
          E. Momjian \inst{12}
          \and
          M. Chiaberge\inst{7,8}
          \and
          B. Trefoloni \inst{13,14}
          \and
          M. Signorini \inst{14,15}
          \and
          X. Ji\inst{3,4}
          \and
          Q. D'Amato \inst{14}
          \and
          G. Risaliti \inst{13,14}
          \and
          R.~D. Baldi\inst{6}
          \and
          A. Fabian\inst{16}
          \and
          H. Übler \inst{17}
          \and
          F. D'Eugenio\inst{3,4}
          \and
          J. Scholtz\inst{3,4}
          \and
          I. Juodžbalis\inst{3,4}
          \and
          M. Mignoli\inst{2}
          \and
          M. Brusa\inst{1,2}
          \and
          E. J. Murphy\inst{9}
          \and
          T.~W.~B. Muxlow \inst{18}
          }

\authorrunning{G. Mazzolari et al.}
   \institute{  Dipartimento di Fisica e Astronomia, Università di Bologna, Via Gobetti 93/2, I-40129 Bologna, Italy
\and INAF – Osservatorio di Astrofisica e Scienza dello Spazio di Bologna, Via Gobetti 93/3, I-40129 Bologna, Italy
\and Kavli Institute for Cosmology, University of Cambridge, Madingley Road, Cambridge, CB3 0HA, UK
\and Cavendish Laboratory, University of Cambridge, 19 JJ Thomson Avenue, Cambridge, CB3 0HE, UK
\and Department of Physics and Astronomy, University College London, Gower Street, London WC1E 6BT, UK
\and INAF - Istituto di Radioastronomia, Via Gobetti 101, I-40129 Bologna, Italy 
\and The William H. Miller III Department of Physics \& Astronomy, Johns Hopkins University, Baltimore, MD, USA
\and Space Telescope Science Institute for the European Space Agency (ESA), ESA Office, 3700 San Martin Drive, Baltimore, MD, USA
\and National Radio Astronomy Observatory, 520 Edgemont Road, Charlottesville, VA 22903, USA
\and Instituto de Radioastronomía y Astrofísica, Universidad Nacional Autónomia de México, Antigua Carretera a Pátzcuaro \# 8701, Ex-Hda. San José de la Huerta, Morelia, Michoacán, C.P 58089, México
\and Max Planck Institut für Astronomie, Königstuhl 17, D-69117, Heidelberg, Germany
\and National Radio Astronomy Observatory, 1011 Lopezville Rd., P.O. Box O, Socorro, NM 87801, USA
\and Dipartimento di Fisica e Astronomia, Università di Firenze, via G. Sansone 1, 50019 Sesto Fiorentino, Firenze, Italy 
\and INAF – Osservatorio Astrofisico di Arcetri, Largo Enrico Fermi 5, I-50125 Firenze, Italy
\and Dipartimento di Matematica e Fisica, Univeristà di Roma 3, Via della Vasca Navale, 84, 00146 Roma RM
\and Institute of Astronomy, University of Cambridge, Madingley Road CB3 0HA, UK
\and Max-Planck-Institut für extraterrestrische Physik (MPE), Gießenbachstraße 1, 85748 Garching, Germany
\and Jodrell Bank Centre for Astrophysics, School of Physics \& Astronomy, The University of Manchester, Alan Turing Building, Oxford Road, Manchester M13 9PL, UK
}
   \date{}


    \abstract
  {The \textit{James Webb Space Telescope} (JWST) has discovered a large population of Active Galactic Nuclei (AGN) in the early Universe. A large fraction of these AGN revealed a significant lack of X-ray emission, with observed X-ray luminosity upper limits 2-3 dex lower than expected.
  We explore the radio emission of these AGN, focusing on the JWST-selected Broad Line AGN (BLAGN, or type 1) in the GOODS-N field, one of the fields with the best combination of deep radio observations and statistics of JWST-selected, spectroscopically confirmed BLAGN. We use deep radio data at different frequencies (144\,MHz, 1.5\,GHz, 3\,GHz, 5.5\,GHz, 10\,GHz), and we find that none of the {37} sources investigated is detected at any of the aforementioned frequencies.
  Similarly, the radio stacking analysis does not reveal any detection down to an rms of ${\sim 0.15}\mu$Jy beam$^{-1}$, corresponding to a $3\sigma$ upper limit at rest frame 5 GHz of $L_{5GHz}=2\times10^{39}$ erg s$^{-1}$ at the mean redshift of the sample $z\sim 5.1$. We compared this and individual sources upper limits with expected radio luminosities estimated assuming different AGN scaling relations, {to check whether these are consistent with the standard BLAGN spectral energy distribution}. For most of the sources the radio luminosity upper limits are still compatible with expectations for radio-quiet (RQ) AGN; nevertheless, the more stringent stacking upper limits and the fact that no detection is found {might suggest} that JWST-selected BLAGN are weaker than standard AGN even at radio frequencies. {Indeed, the probability of having none of the BLAGN detected in none of the investigated radio images is expected to be on average very low ($P<10^{-4}$).}
  We discuss some scenarios that could explain the possible radio weakness, such as free-free absorption from a dense medium, or the lack of either magnetic field or a corona, possibly as a consequence of super-Eddington accretion. These scenarios would also explain the observed X-ray weakness. We also conclude that $\sim$1 dex more sensitive radio observations are needed to better constrain the level of radio emission (or lack thereof) for the bulk of these sources. The Square Kilometer Array Observatory (SKAO) will likely play a crucial role in assessing the properties of this AGN population.}






   \maketitle
%

\section{Introduction}\label{sec:intro}
The \textit{James Webb Space Telescope} (JWST) has discovered a large population of Broad-line Active Galactic Nuclei (BLAGN, or type 1) at $z>4$, identified via the detection of a broad emission component (full-width half maximum $\gtrsim 1000$ km/s ) in the permitted lines (\Ha and/or \Hb). No broad emission is observed in the forbidden lines, such as \OIIIl5007, hence excluding an outflow origin and indicating that the broad permitted lines originate in the Broad Line Regions (BLRs) of AGN \citep[e.g.][]{Harikane23,Maiolino2024_AGNsample,
Kocevski23,ubler23,Taylor2024_AGN}.

This large population of AGN has several peculiar properties. About 10--30\% of them have very red optical colors; therefore dubbed as ``Little Red Dots'' \citep[LRD, e.g.][]{Matthee23,Hainline24,Kocevski2024_LRD}. They are offset from the population of local AGN in classical emission-line diagnostic diagrams, such as the BPT diagrams \citep[e.g.][]{Mazzolari24c,ubler23,Maiolino2024_AGNsample,Harikane23} and lack prominent Fe\,{\sc ii} emission typical of low-redshift AGN \citep{Trefoloni2024_Fe}, possibly due to low metallicity both in the Narrow Line Region (NLR) and in the Broad Line Region (BLR).

One additional, unexpected property characterizing the large majority of these AGN is that their X-ray luminosities are orders of magnitude lower than what is expected assuming the standard BLAGN spectral energy distribution \citep[SED][]{duras20}, thus showing a significant X-ray weakness. \cite{Yue24, Ananna24} and \cite{Maiolino24_X}, analyzing the X-ray properties of $\sim 50$ of these BLAGN found that only very few of them are X-ray detected (1-3\%), despite some of them residing in fields with the deepest X-ray observations ever performed, such as the 7Ms Chandra Deep Field South \cite[CDFS][]{liu17} or the 2Ms Chandra Deep Field North \cite[CDFN][]{xue16}. Even the stack of these objects did not reveal any detection down to X-ray luminosity $\rm L_{2-10keV}\simeq 10^{41}$ erg s$^{-1}$, 
i.e. $\sim 2-3$ dex lower than expected. 
Different scenarios have been invoked to explain this unexpected property.
As outlined in \cite{Maiolino24_X} and \cite{Juodzbalis24_rosetta}, one possibility is X-ray absorption by large, Compton-thick (CTK) hydrogen column densities of dust-poor gas, which would still allow the detection of the broad optical lines. BLR clouds represent an optimal candidate to explain this scenario given their large column densities \citep{Risaliti11, Juodzbalis24_rosetta} and being almost dust-free, provided that their covering factor is much larger ($\sim 4\pi$) than for local AGN (which seems supported by the large equivalent width of the broad component of H$\alpha$). Another possibility is an intrinsic X-ray weakness due to a steep X-ray power spectrum, similar to the one observed in Narrow line Seyfert 1 (NLS1), which, in particular at high redshift, would move most of the X-ray emission out of the observing bands of \textit{Chandra}. Another possible cause of X-ray weakness is the lack of an X-ray-emitting corona, or a much lower efficiency of the corona in producing X-ray photons \citep{Yue24,Maiolino24_X}. \cite{Pacucci24},\cite{Madau2024_SE}, \cite{Lambrides24}, and \cite{King25} have recently proposed that super-Eddington accretion can be at the origin of the observed X-ray weakness; this would also fit well with the similarities with NLS1, given that this class of AGN are thought to be accreting at high rates.

Even if most of the works agreed in classifying these sources as BLAGN, it has also been proposed that the broad Balmer lines observed in these objects, without a counterpart in \OIIIl5007, might not be associated with the BLR of an AGN, but with other extreme phenomena, such as superluminous supernovae, ultra-dense and extremely metal-poor outflows or nuclear compact star-forming cluster, and Raman scattering \citep{Maiolino24_X,Kokubo2024}; however, these alternative explanations fail various observational tests \citep[][Ji et al. in prep.]{Maiolino24_X, Juodzbalis24_rosetta,Inayoshi2024}.

Radio observations may provide important clues on the processes at the origin of the observed properties of this new population of high-z AGN discovered by the JWST, and in particular on their X-ray weakness. AGN radio emission is generally much less affected by obscuration than the X-ray emission \citep{Mazzolari24a, Ricci24} and a high level of radio emission can be produced by processes associated with supermassive black holes (SMBH) activity \citep[e.g. coronal activity, shocks, small-scale radio jets, see][for a review]{panessa19} even in systems that are not classical loud (RL) AGN, as demonstrated by the 'radio-excess' AGN selection techniques \citep{bonzini13,smolcic17b, delvecchio17, whittam22}. Therefore, in the scenario of X-ray weakness due to obscuration, one might expect to detect X-ray weak AGN in the radio band, if the observations are deep enough.

Recently, a few works have investigated the radio properties of high-z AGN or AGN candidates (such as photometrically selected LRD). \cite{Mazzolari24c} investigated the radio counterparts of eight spectroscopically selected JWST BLAGN \citep{Harikane23} over the AEGIS20 radio image \citep{Ivison07}, finding no detection, not even from the stack. The measured radio upper limit was consistent with the typical radio-quiet (RQ) AGN luminosities, excluding the possibility that the sources were instead radio-loud (RL), but it was not possible to provide additional constraints as the radio image was too shallow. 
\citet{Akins24} studied a sample of 434 photometrically selected LRD over the COSMOS-web field \citep{Casey23}, revealing no detection on the COSMOS 3 GHz radio image \citep{smolcic17a} down to $\sim 0.6\mu$Jy at 5$\sigma$ from the stack. Also \cite{Perger24} stacked a sample of 919 LRD  (94\% of them with only a photometric redshift and no spectroscopic confirmation of the AGN nature) over the Very Large Array Sky Survey (VLASS) and the Faint Images of the Radio Sky at Twenty-centimeters (FIRST) surveys. They obtained $3\sigma$ upper limits of $\sim$11$\mu$Jy beam$^{-1}$ and $\sim$18$\mu$Jy beam$^{-1}$ from the stack at 3 GHz and 1.4 GHz, respectively.
Both \cite{Akins24} and \cite{Perger24} find upper limits that are still compatible with expectations for RQ AGN. It is, however, important to note that photometrically selected samples of LRD can be highly contaminated by dusty star-forming galaxies and quiescent galaxies as demonstrated in \cite{PerezGonzalez23}. At the same time, LRDs seem to constitute only a minor fraction of the global high-redshift BLAGN population ($\sim30\%$), given that high-z BLAGN are observed to span a broad distribution of colors and slopes \citep{Hainline24}.

In this work, we aim to investigate the radio properties of spectroscopically confirmed, JWST-selected BLAGN at multiple radio frequencies, in order to provide a uniform and complete characterization of their radio emission. 
Based on available radio images and the location of JWST-selected BLAGN discovered so far, the GOODS-N field provides the best combination of radio sensitivity and BLAGN statistics. Hence, we focus our analysis on this field.

The paper is organized as follows. In Sect.~\ref{sec:data} we present the sample of BLAGN studied in this work and the radio images used for our analysis. In Sect.~\ref{sec:results}, we describe the radio stacking procedure and we present our results in terms of expected radio luminosities versus observed radio luminosities. Finally, in Sect.~\ref{sec:discussion}, we discuss different scenarios that can justify a possible radio weakness of these sources.

Throughout the paper, we assume a flat $\Lambda$CDM universe with $H_{0}=70 \; \rm{km s^{-1} Mpc^{-1}}$, $\Omega_{m}=0.3, \Omega_{\Lambda}=0.7$. 
Unless stated otherwise,  we assume a typical AGN radio spectral index of $\alpha=-0.5$, with $S_{\nu}\propto \nu^{\alpha}$ \citep{Wang24_FP, Bariuan22}. In addition, radio luminosities reported in the form $L_{\nu}$ refer to the AGN rest-frame radio luminosities at frequency $\nu$ in units of erg $\rm s^{-1}$. 

\section{Data} \label{sec:data}

\begin{table*}[!h]
    \centering
    \caption{Sample of BLAGN selected using JWST spectroscopic data on the GOODS-N field.}
    \begin{tabular}{c c c c c c c c c}
   \hline
   \hline
       N & Redshift & $\rm \log (M_{BH}/{M_\odot})$ & $\rm \log (L_{bol}/erg\ s^{-1})$ & $\rm \log (L_{H\alpha}/erg\ s^{-1})$ & $\rm \log (L_{[OIII]}/erg\ s^{-1})$ & Reference\\
       \hline
\\
      
      7 &  [5.08, 5.53] & [7.21, 8.55] & [44.7, 45.8] & [42.38, 43.65] & - & \cite{Matthee23}  \\
      
      10 &  [4.13, 6.82] & [6.3, 7.95] & [43.56, 45.45]& [41.43, 43.32] & [41.73, 43.24] & \cite{Maiolino23c}  \\
      
      1 &  2.26 & 8.47 & 45.65 & 43.53 & 42.65 & \cite{Juodzbalis24_rosetta} (GN-28074)  \\
     
      3 &  2.95, 3.66 & 6.89, 7.04 & 44.07, 43.91 & 41.95, 41.78 & 42.89, 42.53 & \citep{Juodzbalis25}  \\
     
      1 & 6.68 & 8.5 & 44.3 & 42.18 & 42.37 & \cite{Juodzbalis24}  \\
     
      1 & 10.6 & 6.2 & 45 & - & - & \cite{Maiolino23a} (GNz-11)  \\
     {2} & 7.04,7.19 & 7.6,6.7 & 44.35,43.95 & 43.05,42.7  &  43.60, 43.06  & {\cite{Xiao25}} \\
     
     {12} & 3.91,5.2 & 6.85,7.99 & 44.42,44.98 & 42.04,42.68  &  -  & {\cite{Zhang25}} \\
    
     \hline
     
         37  & 5.1 & 7.2  & 44.6 & 42.35 & 42.53 & \\
    \hline

    \end{tabular}
     \tablefoot{In square parenthesis we report the ranges spanned by the sample for each of the physical parameter indicated by the columns.  In the last row we report the median values for each parameter (col. 2-6) and the total number of AGN studied (col. 1). }
    \label{tab:sample}
\end{table*}

\subsection{BLAGN Sample}\label{sec:BLAGN_sample}
In Table~\ref{tab:sample}, we summarize the main physical properties of the {37 BLAGN investigated in this work and collected from the literature}. Despite being drawn from different studies, the selection of these sources is almost uniform. 
Indeed all the sources, except for GNz-11 described in \cite{Maiolino23a} and \cite{Scholtz_2023_GN-z11}, were selected as BLAGN based on the detection of a broad \Ha emission component not revealed in forbidden lines (such as the \OIIIl5007), therefore indicating an emission coming from the BLR.

The bolometric luminosities ($L_{bol}$) of these sources were derived from their broad \Ha luminosities using \cite{Stern12} calibrations.
In addition, all the sources are characterized by Eddington ratios $\lambda_{edd}>0.1$, indicative of significant accretion rates, except for the $z=6.68$ BLAGN presented in \cite{Juodzbalis24} showing $\lambda_{edd}\sim 0.01$.
Different works show that the gas metallicities characterizing these BLAGN are generally well below the solar level \citep[$Z\simeq0.1 Z_{\odot}$,][]{ubler23,Trefoloni2024_Fe,Maiolino24_X}. When available, the estimated stellar masses of their host galaxies range between $\sim \rm 10^{8}-10^{9} M_{\odot}$, and for most of the sources they are significantly undermassive compared to their SMBH mass and local $M_{BH}-M_{*}$ relations \citep{Maiolino23c,Juodzbalis24}.

None of these BLAGN is detected in the 2Ms \textit{Chandra} image of the GOODS-N field, despite being covered by the CDFN observation, which is the second deepest X-ray observation ever performed \citep[with a flux limit of $\sim 10^{-17}$erg cm$^{-2}$ s$^{-1}$, ][]{xue16}. From X-ray stacking analysis \cite{Maiolino24_X}, Comastri et al. in prep. inferred a median observed X-ray luminosity upper limit $\sim 2-3$ dex lower than expected from standard relations between bolometric and X-ray luminosities observed in optically selected and X-ray selected BLAGN \citep{duras20}.

\subsection{GOODS-N Radio images}\label{sec:images}
To investigate the radio properties of these JWST-selected BLAGN, we used a set of radio images at different frequencies available on the GOODS-N field. At 1.5\,GHz we used the radio images produced as part of the first data release of the \textit{e}-MERLIN Galaxy Evolution (\textit{e}-MERGE) Survey \citep{Muxlow20}. These images were obtained by combining $\sim$140\,hrs of observations with the enhanced-Multi-Element Remotely Linked Interferometer Network (\textit{e}-MERLIN) with $\sim$40\,hrs with the Very Large Array (VLA). The e-MERGE DR1 released three 1.5\,GHz images characterized by different beam-size sensitivity combinations, but all covering a $\sim 15'\times 15'$ field of view.
Among these, we considered the image with the best resolution ($0.28\times 0.26$ arcsec$^2$), which also has the best point-source sensitivity, 
reaching an rms noise of $1.17 \mu$Jy beam$^{-1}$.

The 3\,GHz radio image (Jimenez-Andrade et al. in prep.) was obtained with 64 hours of VLA observations in the A configuration using the S-band (2-4 GHz) receivers (project ID VLA/22A-317; PI Jiménez-Andrade). Two pointings were needed to provide a nearly homogeneous sensitivity across the GOODS-N field. The point-source sensitivity of the mosaic is 0.76 $\mu$Jy/beam and the angular resolution is $0.62\times0.62$ arcsec$^2$. 

We also considered higher frequency deep radio coverage of the GOODS-N field, in particular, the 5.5\,GHz \citep{Guidetti17} and 10\,GHz \citep{Jimenez24} surveys, both taken with the VLA, reaching sensitivities of $\sim 3\mu$Jy  beam$^{-1}$ and $\sim 0.7 \mu$Jy beam$^{-1}$, and resolutions of 0.5 arcsec and 0.22 arcsec, respectively. 

Finally, we considered the second data release of the 144 MHz 
LOw-Frequency ARray (LOFAR) Two-metre Sky Survey (LoTSS; \citealt{Shimwell22}), which includes the GOODS-N field in the surveyed area. The LoTSS survey image has an angular resolution of 6 arcsec and a median rms in the region covered by our sources of $\sim 70 \; \mu$Jy  beam$^{-1}$. While LOFAR is quite poor in spatial resolution, the sub-arcsec resolution of the 1-10 GHz images implies we are probing physical scales of 1-4 kpc at the median redshift of the sample.  

Covering a wide range of frequencies allows us to account for different shapes of the radio spectra and enhances our chances to get a detection of the sources.
Assuming typical optically thin synchrotron spectra (radio spectral index $\alpha$=-0.7), the 1.5\,GHz and 3\,GHz images are those that provide the best point source sensitivity. Assuming an optically thick flat spectrum population ($\alpha$=0) the 3\,GHz and 10\,GHz images are the most sensitive ones. In the following, we will use the results obtained at 3\,GHz as a reference for the analysis, {and we will derive 5\,GHz rest frame radio luminosities,} assuming a typical AGN spectral index of $\alpha$=-0.5.

\section{Methods and Results}\label{sec:results}

\begin{figure}[!h]
    \centering
    \includegraphics[width=1\columnwidth]{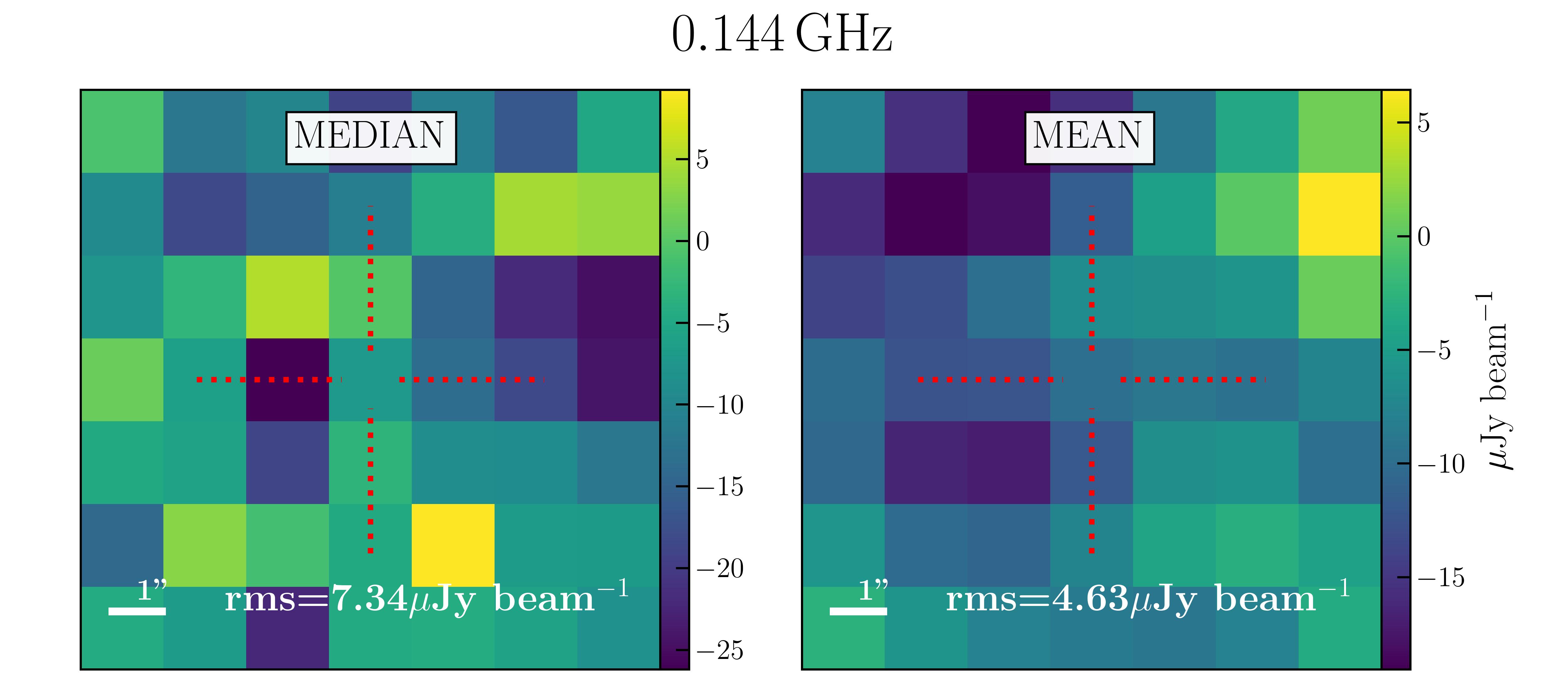}
    \includegraphics[width=1\columnwidth]{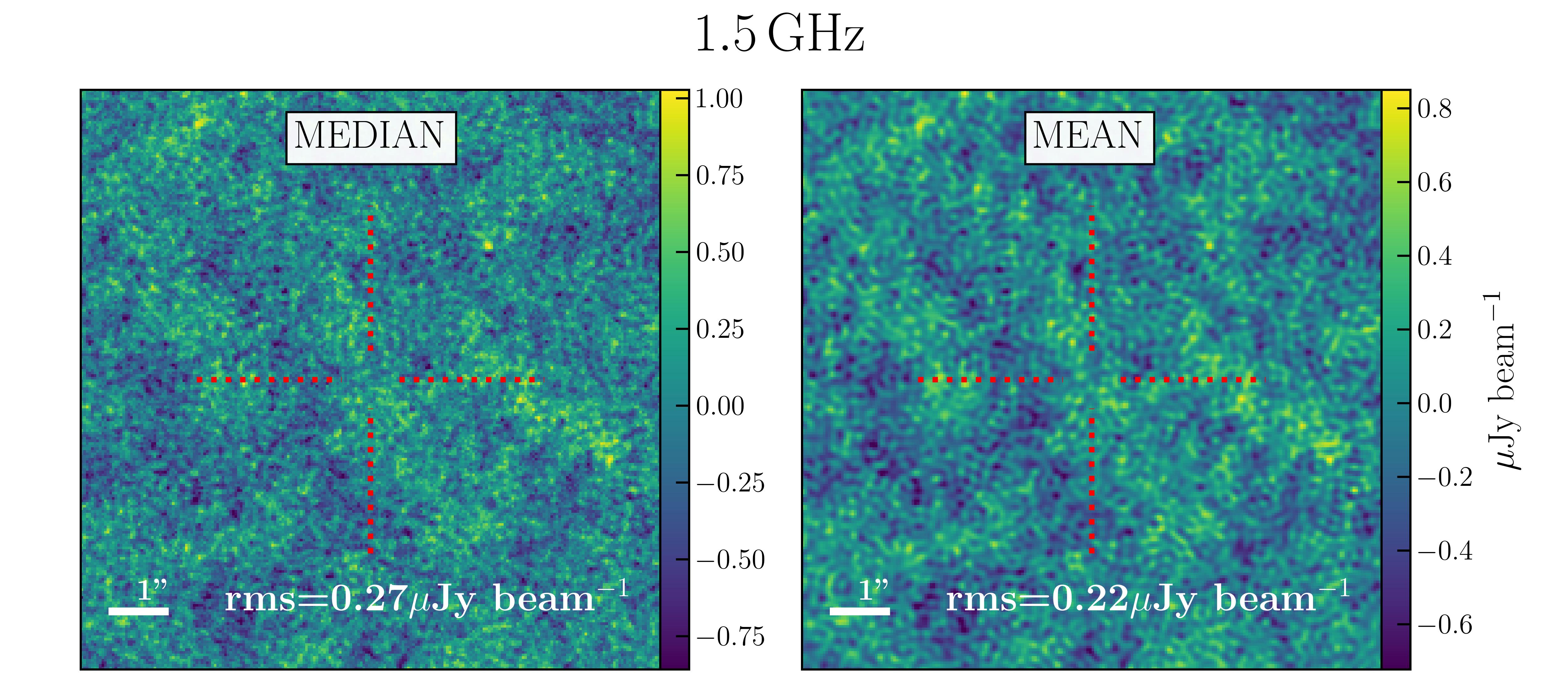}
    \includegraphics[width=1\columnwidth]{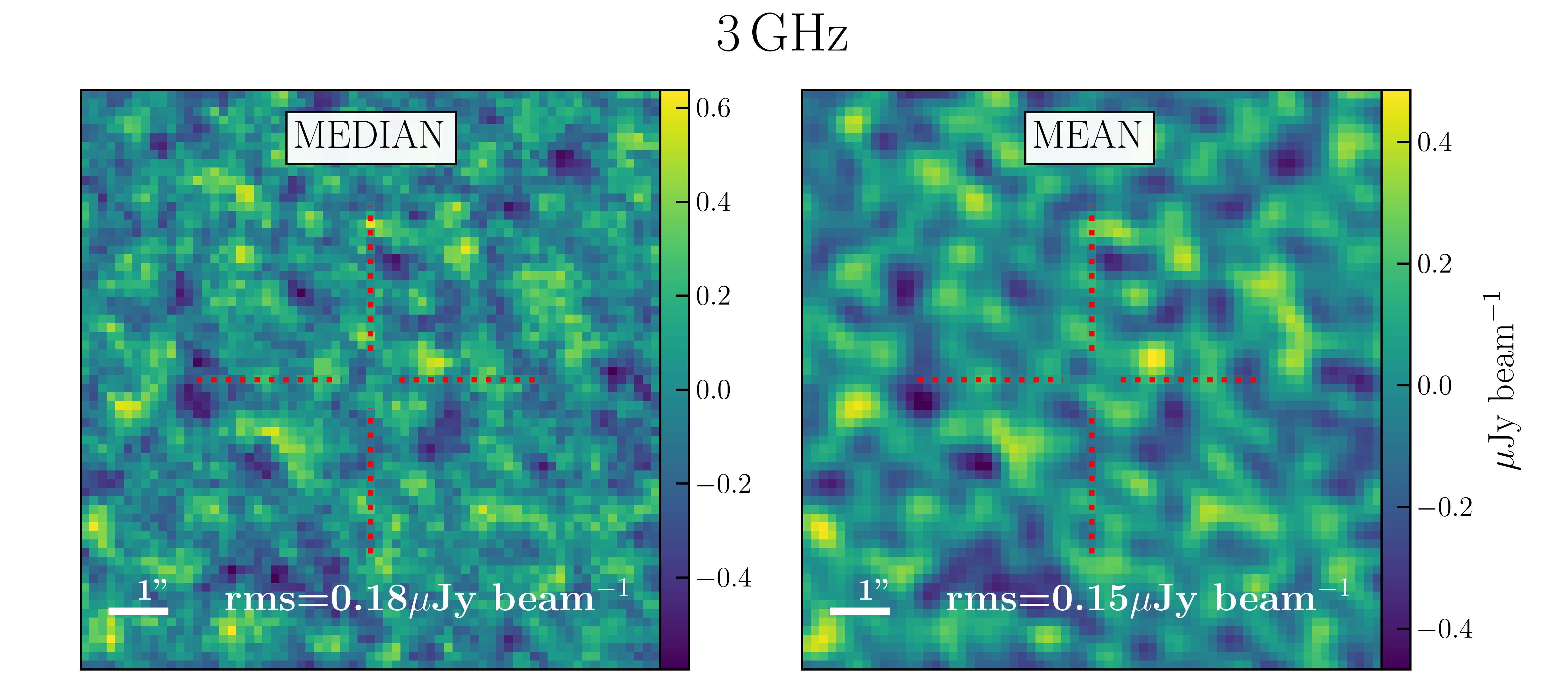}
    \includegraphics[width=1\columnwidth]{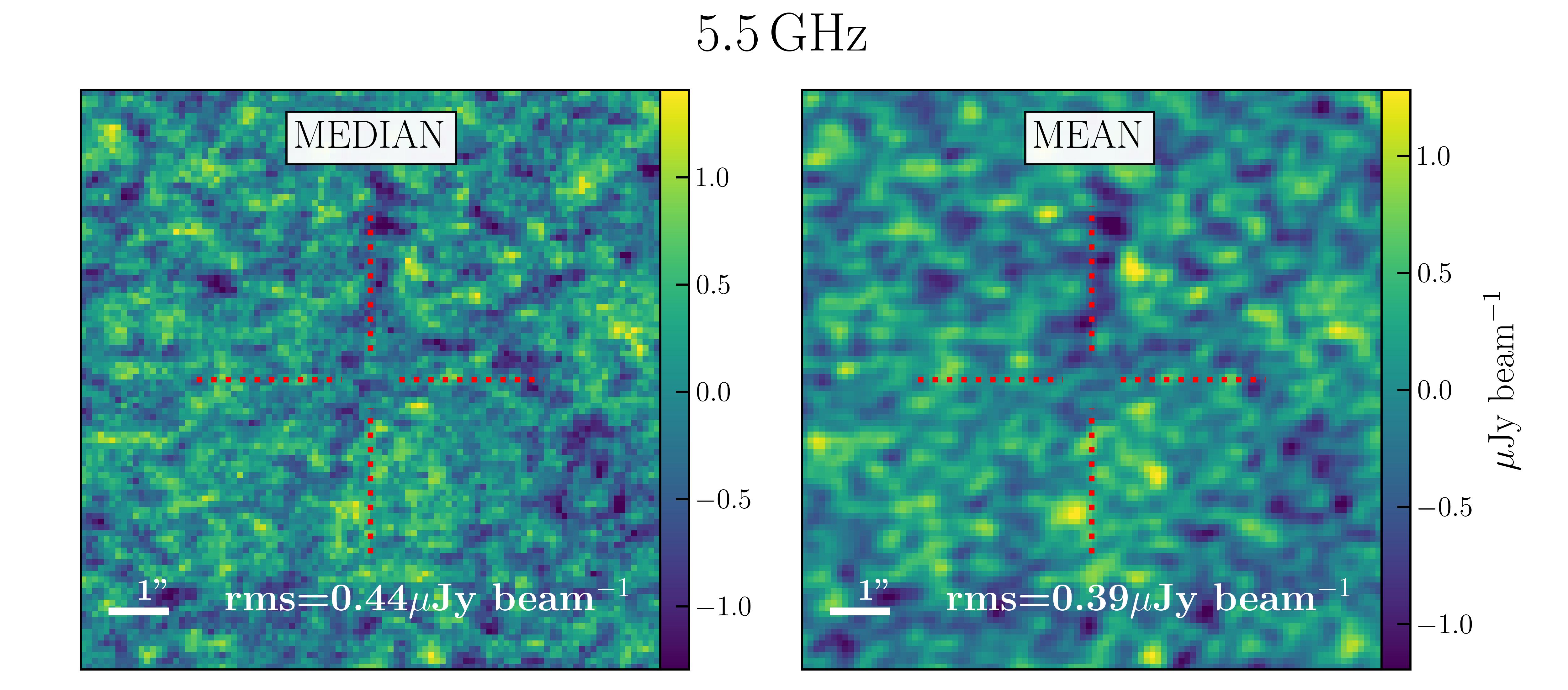}
    \includegraphics[width=1\columnwidth]{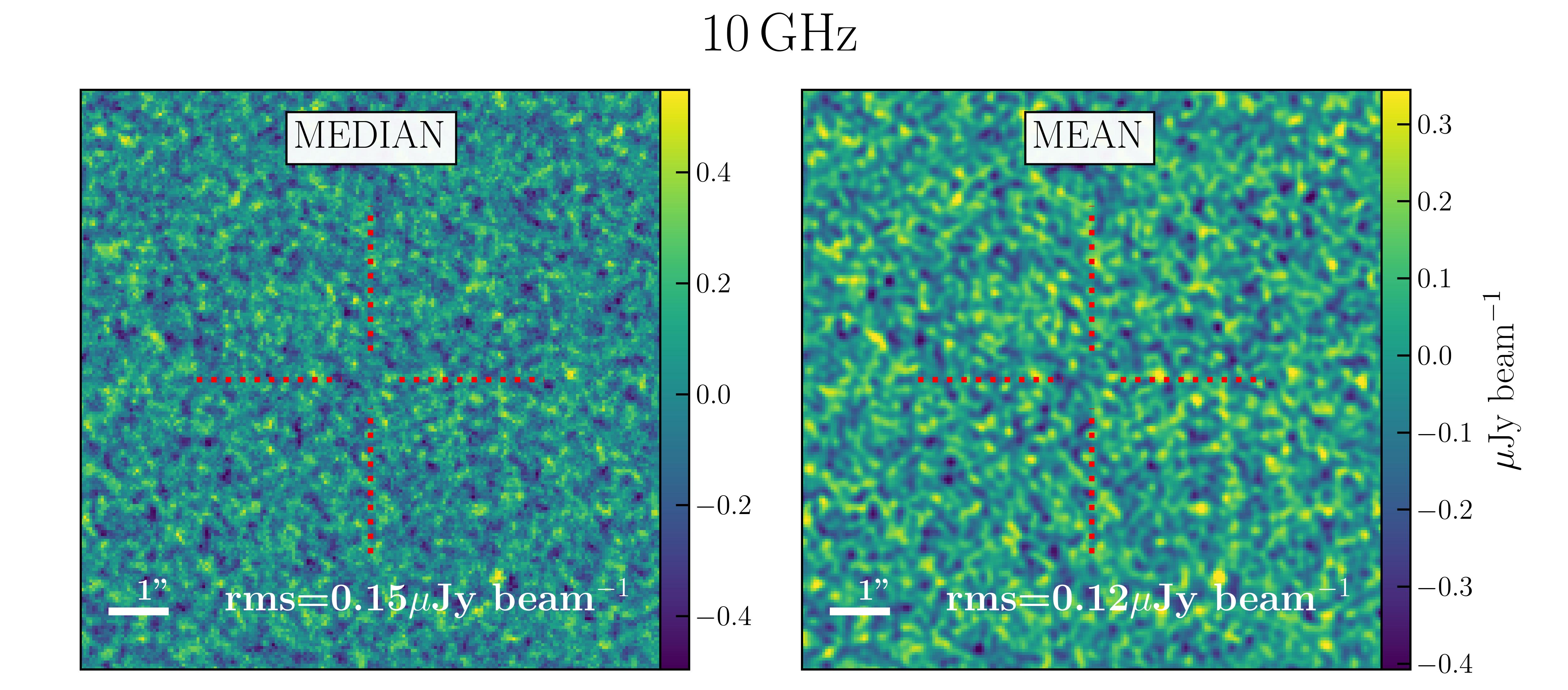}
    \caption{$10"\times 10"$ cutouts of the mean (right) and median (left) radio stacking of the {37 JWST selected BLAGN} on the GOODS-N field performed on the images presented in Sect.~\ref{sec:images}. 
    The values of the rms measured in the stacked images are reported in the panels. }
    \label{fig:rstack}
\end{figure}

\subsection{Radio stacking}\label{sec:stacking}
All the 37 BLAGN fall inside the aforementioned radio images, but none of them is detected down to a 3$\sigma$ level. We hence performed a radio stacking analysis to search for indications of under-threshold radio emission.

The radio stacking was performed as follows. For each of the 37 sources, we made a $10"\times 10"$ (or $25"\times 25"$ for the LOFAR images) cutout centered on the position of the source. These images were then aligned, performing a pixel-by-pixel stacking considering both a median and a mean stack. We did not consider a weighted mean procedure since all the sources lie in regions of the radio images where the rms noise is fairly uniform. The stacking was performed for all the five radio images described in Sect.~\ref{sec:images}, but in none of them we detected a significant (3$\sigma$ level) radio emission. The rms noise and the peak brightness densities of each stack are reported in Table~\ref{tab:stack}. In Fig.~\ref{fig:rstack} we show both the median and mean stack performed on all the images investigated in this work. In particular, for the 3\,GHz radio image, our reference image, we reached an rms noise of $\sim 0.2\,\mu$Jy\,beam$^{-1}$. This corresponds to a $3\sigma$ upper limit of $\log L_{3GHz}=39.3$ erg s$^{-1}$ or to a 5 GHz rest-frame radio luminosity 
of $\log L_{5GHz}=39.4$ erg s$^{-1}$ at the median redshift of the sources ($z\sim5.2$).

\begin{table}[!h]
    \centering
    \caption{The rms noise and peak brightness density of the stacked images of the 37 BLAGN performed using the radio images described in Sect.~\ref{sec:images}}
    \begin{tabular}{c | c c | c c }
\hline
\hline
        $\nu_{obs}$ [GHz] & \multicolumn{2}{c|}{rms [$\mu$Jy beam$^{-1}$]} & \multicolumn{2}{c}{$f_{peak}$ [$\mu$Jy beam$^{-1}$]} \\
        & Mean & Median & Mean & Median \\
        \hline
        0.144 & 4.63 & 7.34 & -5.92 & 5.14 \\
        
        1.5 & 0.22 & 0.27 & 0.18 & 0.23 \\
        
        3   & 0.15 & 0.18 & 0.15 & 0.18 \\

        5.5 & 0.39 & 0.44 & 0.11 & 0.56 \\

        10  & 0.12 & 0.15 & 0.05 & 0.13 \\
\hline

    \end{tabular}
    \tablefoot{Peak brightness are computed considering the maximum intensity in the $3\times3$ pixels (the pixel size of the 3\,GHz image is 0.15") region around the central one.}
    \label{tab:stack}
\end{table}

\begin{figure*}[h!]
    \centering
    \includegraphics[width=0.8\linewidth]{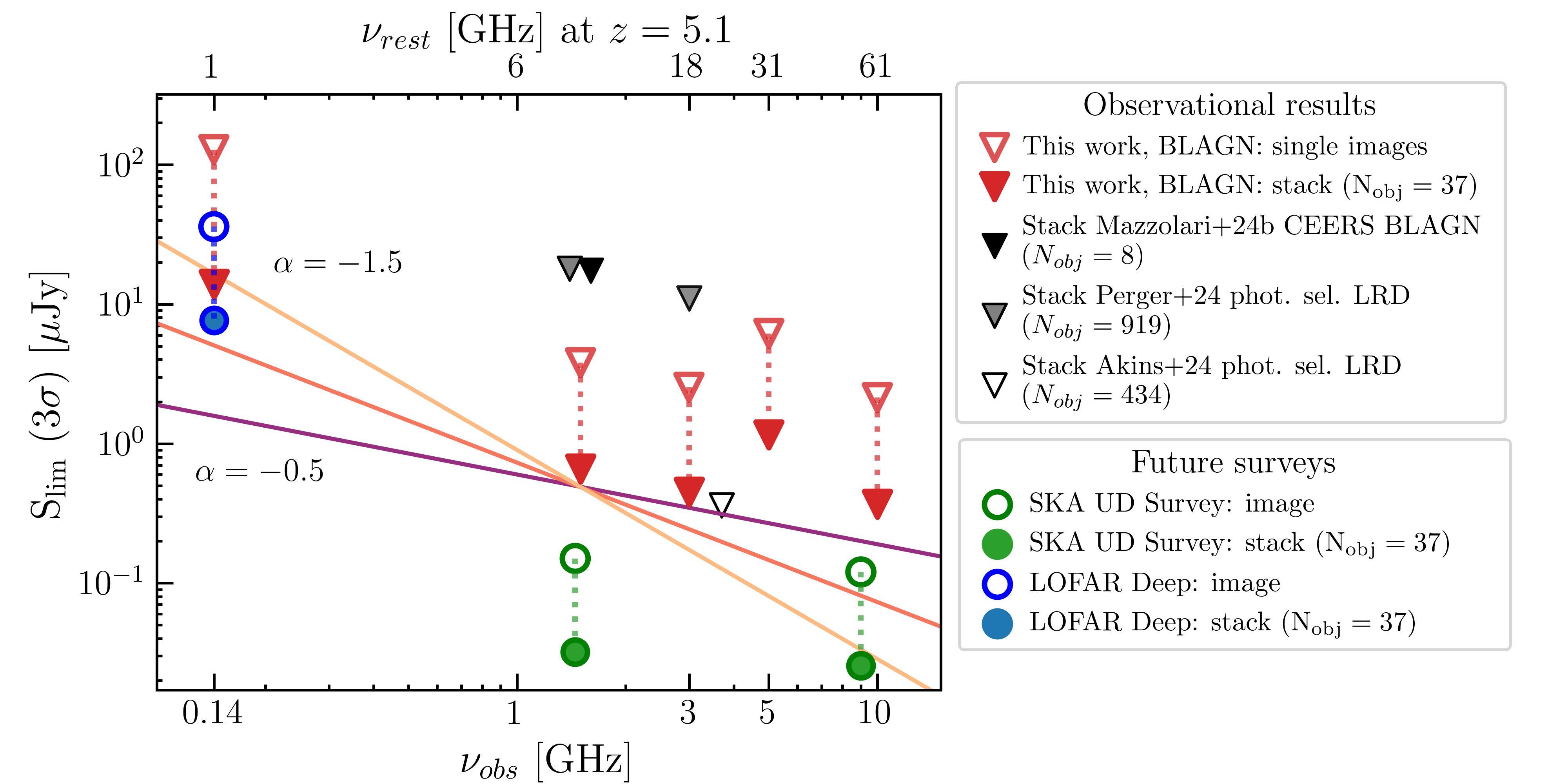}
    \caption{Distribution of the radio $3\sigma$ upper limits obtained at the frequencies of the images described in Sect.~\ref{sec:images}. With empty red triangles and filled red triangles we show the sensitivities of the radio images and of the stack, respectively. The green empty circles refer to the sensitivities of the SKAO UD radio survey at $\sim 1$ GHz and $\sim 10$ GHz. The blue empty circle refers to the LOFAR deep radio image of the GOODS-N field. Filled blue and green circles refer to the sensitivities that would be reached considering the stack of the 37 BLAGN investigated in this work on the deep LOFAR and UD SKAO images, to be compared with the stack achieved in the currently available images. The black triangle refers to the radio stack of the 8 high-z BLAGN performed in \cite{Mazzolari24c} on the AEGIS20 1.4GHz image \citep{Ivison07}. Instead, the gray and empty tringles refer to the radio stacks of large samples of photometrically selected LRD performed in \cite{Perger24} and \cite{Akins24}. }
    \label{fig:sensitivity}
\end{figure*}

In Fig.~\ref{fig:sensitivity} we show the upper limits distribution obtained from the stacks of the different images. We also show the upper limits derived by the stacks performed in previous studies (see Sect.\ref{sec:intro}). In particular, the stacked images analysed in the present work are $> 10$ times deeper than those in \cite{Perger24} (919 photometrically selected LRD) at comparable frequencies. The 3 GHz stack performed in \cite{Akins24} (434 photometrically selected LRD) reaches a similar depth to our 3 GHz stack. However, we remark that photometric selections of LRD likely provide largely contaminated samples, as shown in \cite{PerezGonzalez23}, and therefore they are not ideal to trace the average properties of BLAGN, like those investigated in this work, which are all spectroscopically confirmed. On the contrary, the radio upper limit derived by \citet{Mazzolari24c} was obtained considering the 8 BLAGN spectroscopically selected by \cite{Harikane23} among the medium resolution (MR) spectra of the CEERS survey \citep{Finkelstein22}, i.e. belonging to the same BLAGN population studied here. However, the constraints derived by \cite{Mazzolari24c} are much shallower than those obtained in this work due to the limited sample and the shallower available radio image ($\sim 10\; \mu$Jy rms).

In Fig.~\ref{fig:sensitivity}, we also show the sensitivities of the ongoing deep LOFAR imaging of the GOODS-N field ($\sim 200$ hrs of observations, likely reaching rms $\sim 12 \; \mu$Jy; Vacca et al. in prep.), and of the reference Square Kilometre Array (SKA) radio-continuum surveys. In particular we show the SKA survey ultra-deep (UD) tiers at $\sim 1$ GHz and $\sim 10$ GHz, respectively reaching sensitivities of $\sim 0.05\; \mu$Jy and $\sim 0.04\; \mu$Jy over areas of $\rm \sim 1$ deg$^{2}$ and $\sim 0.01$ deg$^{2}$  \citep{prandoni15}. 

In the plot, we also show the sensitivities that would be reached considering the stack of our 37 BLAGN using the future deep LOFAR and UD SKA surveys. Clearly, when SKAO will become fully operational, we will expect much larger samples of JWST-selected BLAGN, given the much larger area that will be covered by these radio surveys, hence these data points obtained considering only 37 objects are just for comparison with our stacked sample. As it is possible to see, the SKA UD surveys will reach a factor of $\sim 20$ deeper sensitivities than currently available at $\sim 1$ GHz and $\sim 10$ GHz, while the GOODS-N LOFAR deep image will be $\sim 5$ times deeper than the currently available LoTSS survey.

The distribution of the stack upper limits obtained with the radio images considered in this work, clearly shows that, in the optimistic case of a source with a flux density just below the observed 1.5\,GHz $3\sigma$ sensitivity, this source would be detectable in the currently available LOFAR image only if characterized by an ultra-steep spectral index $\alpha\lesssim -1.3$. The constraint on the spectral index would relax to $\alpha\lesssim -1$, when the LOFAR deep observations of the GOODS-N will be released.

\subsection{Expected radio luminosities}\label{sec:Lradexp} 
To verify how the radio upper limits that we obtained for these BLAGN (both for the individual sources and for the stack) compare with the AGN expectations, we have to estimate their expected radio luminosities under the assumption that these sources are 'standard' AGN, i.e. that they can be described by standard BLAGN SED in terms of optical, X-rays, and radio emission \citep{Richards06, Shang11, liu21}. {Our goal is to test the null hypothesis that the radio emission of these sources is consistent with what is predicted by the standard SED of BLAGN.}

There are different scaling relations that can be exploited to estimate the expected radio luminosity for these BLAGN, given the available observational quantities reported in Table~\ref{tab:sample}. 
In particular, we focus on scaling relations observed for RQ AGN populations, as the luminosity upper limits that we obtain are too faint for a RL AGN population (as demonstrated in Appendix ~\ref{sec:RL}).

Some of the most thoroughly studied AGN relations involving radio emission are the ones between X-ray and radio luminosity \citep{damato22,panessa15,Fan16}, as well as the so-called Fundamental Plane (FP) relations \citep{merloni03, Wang24_FP, Bariuan22}. 
For both these kinds of relations, given the intrinsic AGN X-ray luminosity, and for the FP relations also the SMBH mass, they allow the estimation of the expected radio luminosity. All the JWST-selected BLAGN are undetected in X-rays. {However, as reported in Sect.~\ref{sec:intro}, we still do not know whether this is because X-ray photons are produced but are absorbed by a very dense gaseous environment or if it is due to an intrinsic X-ray weakness. We assumed (and therefore tested) the scenario in which the intrinsic X-ray emission of our sources is consistent with the standard relations and its observed weakness is due to absorption.} Therefore, we computed their expected intrinsic X-ray luminosities {using the direct relation between the luminosity of the broad \Ha emission and the 2-10 keV luminosity \citep{Jin12}}. As mentioned above, we are assuming the scenario whereby the intrinsic X-ray and radio emission follow the standard type 1 AGN SED, and we want to test if this hypothesis is compatible with radio non-detection. { Only in this way we will prove whether the JWST-discovered AGN are characterized by a "non standard" emission even in the radio band.}
Given the intrinsic $L_{\rm 2-10keV}$, the expected radio emission is computed considering the X-ray to radio luminosity scaling relation derived by \cite{damato22} and the FP relation derived by \cite{Bariuan22}. We chose these relations because they were derived considering RQ AGN samples selected from some of the deepest fields in terms of radio and X-ray flux sensitivities \citep[especially the X-ray to radio luminosity relation of \citealt{damato22}, see also][]{Mazzolari24a}. In addition, such samples span a large range in redshift, including sources at redshifts comparable to those of the BLAGN studied in this work \citep[the FP relation of ][was obtained from sources extending up to z$\sim 5$]{Bariuan22}. Furthermore, these relations were derived without pre-selecting sources with a specific kind of accretion physics, preventing us from making any a priori assumption on the accretion mechanism of the JWST-selected BLAGN.

Another approach one can follow is to use the well-known relation between the radio luminosity and the dust-corrected \OIIIl5007 luminosity \citep{Xu99,DeVries07,Berton16,baldi19,Baldi21} observed for both RQ and RL AGN. Here, we considered the one derived in \cite{DeVries07}, considering RQ sources including both bright BLAGN and low radio luminosity AGN from the Sloan Digital Sky Survey (SDSS). The advantage of this approach is that we can directly use the observed \OIIIl5007 luminosities with no need to pass through other quantities. Unfortunately, \OIIIl5007 luminosities are only available for 15 out of the 37 BLAGN; for the other sources the \OIIIl5007 falls outside of the JWST spectral coverage or in a spectral gap. {While it is possible that the \OIIIl5007 line might be contaminated by host galaxy emission, we will show in Sect.~\ref{sec:sfr} that the expected radio luminosity coming from star formation (SF) in these sources is $>1$ order of magnitude lower than the AGN one, therefore it is not expected to significantly contaminate also the \OIIIl5007 line. }

Finally, the radio emission can be estimated by considering the radio loudness parameter $ R$, defined as the ratio between the AGN radio luminosity density at rest-frame 5\,GHz (i.e. in erg $\rm s^{-1}\,Hz^{-1}$) and the optical luminosity density (in the rest-frame B band)  \citep{Kellermann1989}. Different works investigated the intrinsic distribution of the parameter $R$ for optically selected RQ AGN, both from an observational and theoretical perspective \citep{Arnaudova24,white07,Cirasuolo03, Balokovic12}. There is no general consensus on which value of $R$ corresponds to the peak (or mean) of the RQ AGN distribution, but the aforementioned works overall agree in setting it between $0<\log R<1$, while for $\log R>1-1.5$ AGN are generally classified as RL \citep{Bariuan22, Kellermann1989}. Therefore, for our analysis we considered two distinct fiducial values of the radio loudness parameter: $\log R=0$ and $\log R=1$.
Regarding the optical luminosity in the B-band, we cannot simply take the observed value, as for these intermediate luminosity AGN the continuum is generally contributed significantly by the host galaxy's stellar light \citep{Maiolino2024_AGNsample,Juodzbalis24}. Therefore, we used the luminosity of the broad component of the \Ha line, which is entirely associated with the AGN, and from this we derived the AGN optical luminosity in the rest-frame B band using the scaling relation between the H$\alpha$ broad emission line and B-band given by \cite{Stern12}. Then we used the assumed $R$ to derive the expected radio luminosity.

\subsection{Expected vs observed radio luminosities}\label{sec:radio_weakness}

\begin{figure*}[h!]
    \centering
    \includegraphics[width=0.98\linewidth]{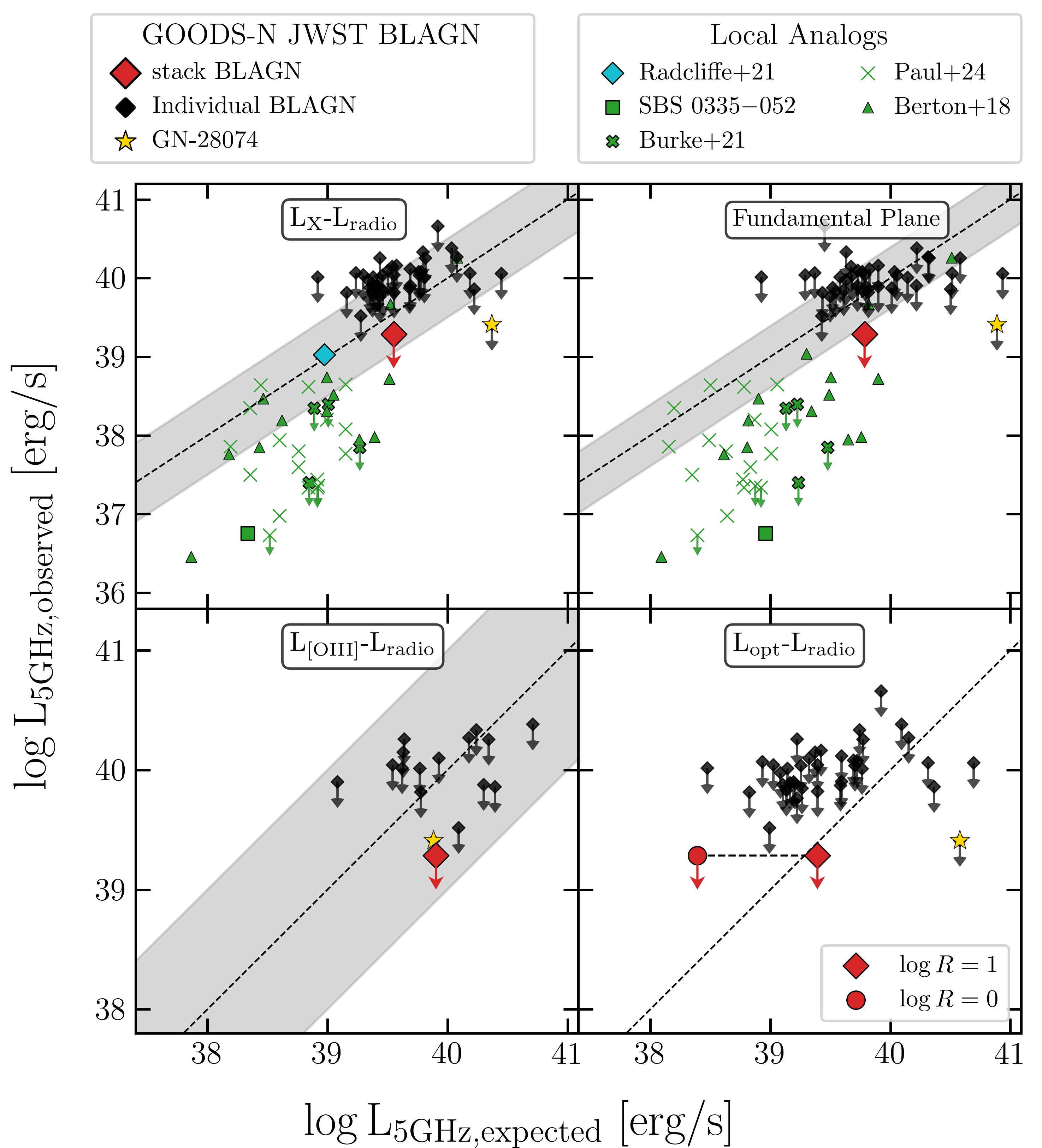}
    \caption{Observed rest-frame 5GHz radio luminosity versus expected rest-frame 5GHz radio luminosity of different samples of high-z and local BLAGN according to the four RQ AGN relations described in Sect.~\ref{sec:Lradexp}. The black data points represent the JWST-detected BLAGN analyzed in this work, the red diamond their stack, the gold star the BLAGN GN-28074 reported in \cite{Juodzbalis24_rosetta}. The green symbols indicate sources from different local analog samples. The green square represents the position of SBS 0355-052 \citep{Hatano24,Johnson09}, the thick crosses represent the metal-poor dwarf BLAGN reported in \cite{Burke21}, the thin crosses the X-ray weak BLAGN reported in \cite{Paul24}, the filled triangles the radio detected and RQ NLS1 reported in \cite{Berton18}. The light blue data point in the top-left panel represents the stack of the X-ray selected but radio undetected AGN on the GOODS-N field performed by \cite{Radcliffe21_rstack}. The red diamond and the black data points in the bottom-right panel, are computed considering $\log R=1$, while the red circle indicate the expected radio luminosity assuming $\log R=0$. The gray shaded areas represent the scatter of the relations, while the black dashed line is the 1:1 relation.}
    \label{fig:RQ_panel}
\end{figure*}

In Fig.~\ref{fig:RQ_panel} we show the observed radio luminosities' upper limits (derived at $3\sigma$) compared with the rest frame $5$\,GHz expected radio luminosities from various scaling relations. More specifically, the four panels refer to the four different relations used to estimate the expected radio luminosities presented in Sect.~\ref{sec:Lradexp}. {Similarly, in Appenidx~\ref{app:RQ_obs} and in Fig.~\ref{fig:RQ_panel_referee}, we present the same plot but showing the observed radio luminosity upper limits directly compared to the observed quantities from which the expected radio luminosities are computed in Sect.~\ref{sec:Lradexp}:  the broad \Ha luminosity, M$_{BH}$, the \OIIIl5007 luminosity and the 4400Å luminosity. }

While in general most of the individual BLAGN have $3\sigma$ radio upper limit above the expected radio luminosities 
the upper limit corresponding to the stack lies below the 1:1 line in all the panels (except for the $\rm L_{opt}-L_{radio}$ relation in the case $R=0$), even if it is still compatible with the scatter of the relations. In some cases also the individual upper limits lie significantly below the 1:1 lines. Most remarkably, GN-28074 (the so called 'Rosetta stone' presented in \citealt{Juodzbalis24_rosetta}), shows a radio upper limit significantly below its expected radio luminosity, with a deviation of $\sim 1$ dex in all panels, much larger than the scatters of the relations. 

For comparison, in the two upper panels of Fig.~\ref{fig:RQ_panel}, we also show the positions of various local AGN, whose physical properties (in terms of metallicities, X-ray deficit, host galaxy properties or SMBH mass) are similar to the ones of the high-$z$ AGN discovered by the JWST and studied in this work (see Sect.~\ref{sec:BLAGN_sample}). These sources (green markers) are as detailed below:
\begin{itemize}
    \item 
SBS 0335-052 (green square) is a low metallicity ($\sim 1/40 Z_{\odot}$) dwarf galaxy, initially thought to be a pure star-forming galaxy (SFG), that was recently found to host an AGN, based on the detection of broad \Ha emission lines, high ionization lines, and mid-IR variability \citep{Hatano24,Hatano2024b}. Despite these features, it remains undetected in X-ray, with only a weak signal attributed to its host galaxy. Its X-ray weakness is as extreme as the one measured for GN-28074 \citep{Juodzbalis24_rosetta}. 
\item Four spectroscopically-selected BLAGN from the SDSS, that are hosted in metal-poor dwarf galaxies (\citealt{Burke21}; thick green crosses). Like the sources investigated in this work, these AGN are metal poor and all undetected in X-ray and show X-ray luminosity upper limits that are 1-2 dex lower than expected, assuming they are standard BLAGN.  
\item 
The X-ray weak local AGNs studied in the radio band by \citet[][green crosses]{Paul24}. These AGN exhibit significant X-ray weakness, based on their X-ray-to-UV luminosity ratio, but do not show significant obscuration from their X-ray spectral analysis ($\log N_H<22$). These sources are also generally characterized by high accretion rates ($\lambda_{\rm edd}>0.1$).
\item 
the 23 radio-detected NLS1 galaxies (green triangles) presented by \cite{Berton18}. NLS1 are a class of AGNs with broad permitted lines, which are narrower than in standard BLAGN (but still broader than the forbidden lines), implying SMBH with masses $10^5-10^8 M_{\odot}$. In addition, they are often characterized by high accretion rates, approaching the Eddington limit. NLS1 show steep X-ray spectra \citep[with photon indices that can approach $\Gamma\sim3$ (compared to the usual value of $\Gamma\sim 1.9$;][]{Vaughan99}, resulting in a "natural" X-ray weakness in the harder part of the X-ray spectrum. The analogy with NLS1 has been suggested by \cite{Maiolino24_X} as one of the possible (but not exhaustive) solutions to address the origin of the X-ray weakness of the high-z JWST-selected BLAGN. 
\end{itemize}
For all these local analogs we took publicly available rest frame $\rm L_{5GHz}$, and when the radio luminosity was provided at a different rest frame frequency we converted it into rest frame $\rm L_{5GHz}$ using the observed radio spectral index or $\alpha=-0.5$ when the spectral index was not available (the same $\alpha$ assumed for our BLAGN analysis).\\
Most of the aforementioned local analogs show radio luminosities (or radio upper limits) that are 0.5-1.5 dex lower than their expected values, based on the considered scaling relations, suggesting that the observed X-ray weakness may be associated with an observed radio-weakness. The physical properties of these local sources (including the X-ray weakness) are similar to those of the JWST detected BLAGN,  suggesting that also the latter might show a "radio-weak" behavior.

In the first panel of Fig.~\ref{fig:RQ_panel} we also plot the position of the radio stack performed by \cite{Radcliffe21_rstack} obtained from 88 X-ray detected AGN in the GOODS-N field that are undetected in the deep 1.5\,GHz e-MERGE radio image. The radio stack returned a clear detection (S/N$\sim 10$).
Interestingly, the stacked radio luminosity of these AGN nicely falls on the \cite{damato22} relation, suggesting this population might be dominated by standard RQ AGN. Contrarily, the JWST-selected BLAGN investigated in this work are not detected in X-rays (showing a significant X-ray weakness), and are not detected in the radio stack, suggesting that they probably constitute a different population of AGN compared to the sources analyzed in \cite{Radcliffe21_rstack}.

\section{Discussion}\label{sec:discussion}

\subsection{Radio emission from star formation}\label{sec:sfr}
\begin{figure}[h!]
    \centering
    \includegraphics[width=1\linewidth]{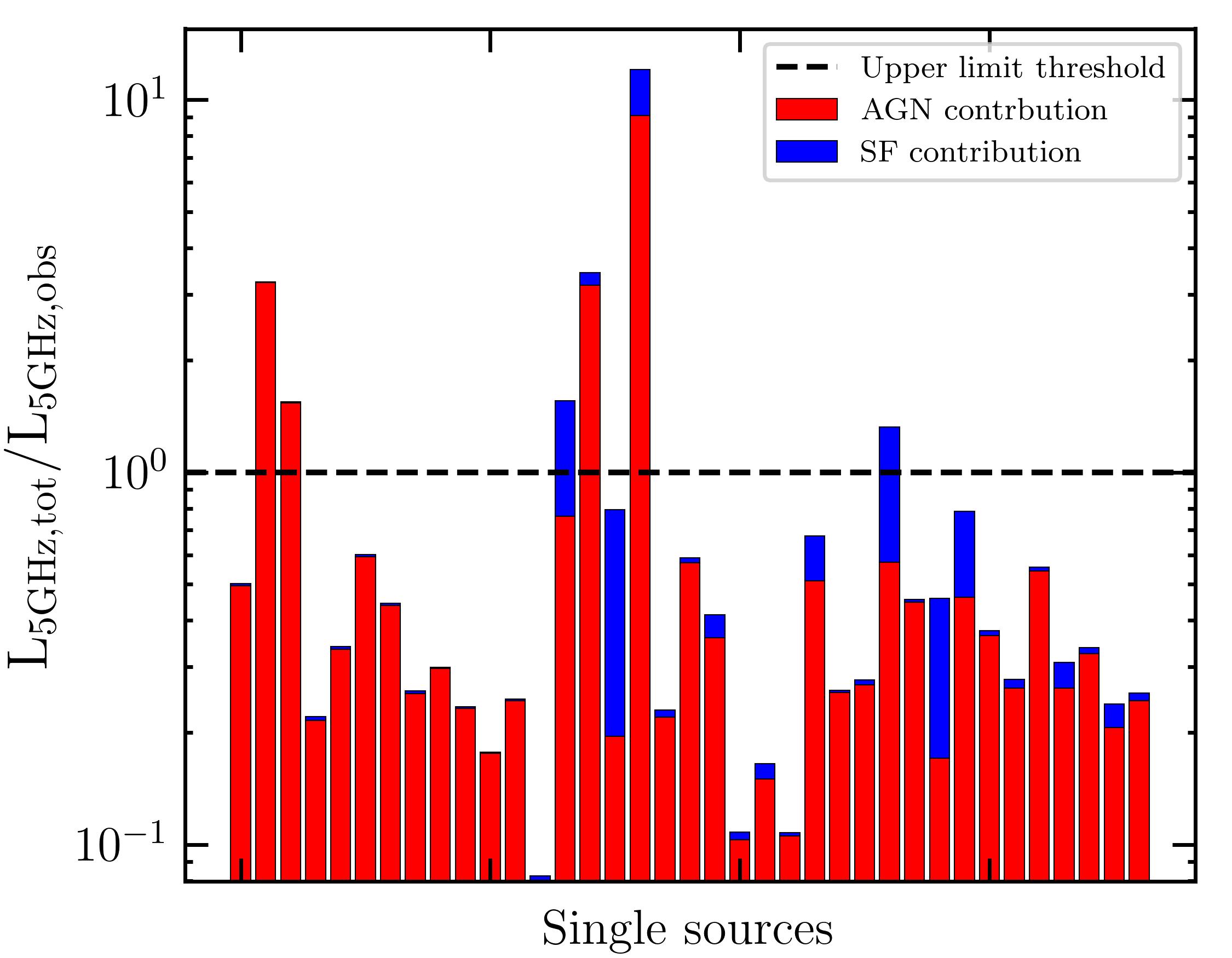}
    \caption{{Distribution of the total 5 GHz radio luminosities ($\rm L_{5GHz,AGN} + L_{5GHz,SFR_{MS}}$) of the JWST selected BLAGN investigated in this work normalized by the 3$\sigma$ 5 GHz luminosity upper limit obtained for each of these sources. For each source we show the fractional contribution to the total radio luminosity of the AGN and SF, as described in Sect.~\ref{sec:sfr}. Sources crossing the black dashed line ($\rm L_{5GHz, tot}= L_{5GHz, obs}$) are those in tension with the radio undetection.}}
    \label{fig:AGN_vs_SFR}
\end{figure}

The observed radio emission of the JWST selected BLAGN might not only be determined by their AGN contribution but also by the contribution of SF. The star formation rates (SFRs) of the sources investigated in this work are highly uncertain due to the difficulties in decomposing the AGN from the host-galaxy emission via SED fitting. To provide an estimate of the amount of radio emission produced by SF in these sources, we proceeded as follows. We took the available stellar mass \citep[for sources from][]{Maiolino23c, Juodzbalis24_rosetta, Juodzbalis24,Juodzbalis25}, and we considered as a reference value for the other sources the average stellar mass of the sample, that is $\log M_{*}=8.36$ M$_{\odot}$.
Then, we computed for each source the main-sequence (MS) SFR using the relation from \cite{Popesso23} and finally, using \cite{novak17}, the expected SF-related radio emission. {In Fig.~\ref{fig:AGN_vs_SFR} we show for each of the sources studied in this work the fractional contribution of AGN and SF to the total radio emission (normalized to the observed radio luminosity upper limit). In particular, the AGN-related radio emission corresponds to the median value of the different expected AGN radio luminosities computed in Sect.~\ref{sec:Lradexp}. For the vast majority of the sources $\rm L_{5GHz,AGN}$ is much larger (10-100 times) than $\rm L_{5GHz,SFR_{MS}}$, and the contribution of SF to the global radio luminosity is negligible. However, there are also a few cases where the two contributions are comparable and where the sum of the AGN and SF radio emission crosses the threshold set by the observed radio luminosity upper limit. If these sources really host MS galaxies, this analysis further supports the scenario in which the AGN-related radio emission must be weaker. We will discuss more this point in Sect.\ref{sec:RQonly}}.

\begin{figure}[h!]
    \centering
    \includegraphics[width=1\linewidth]{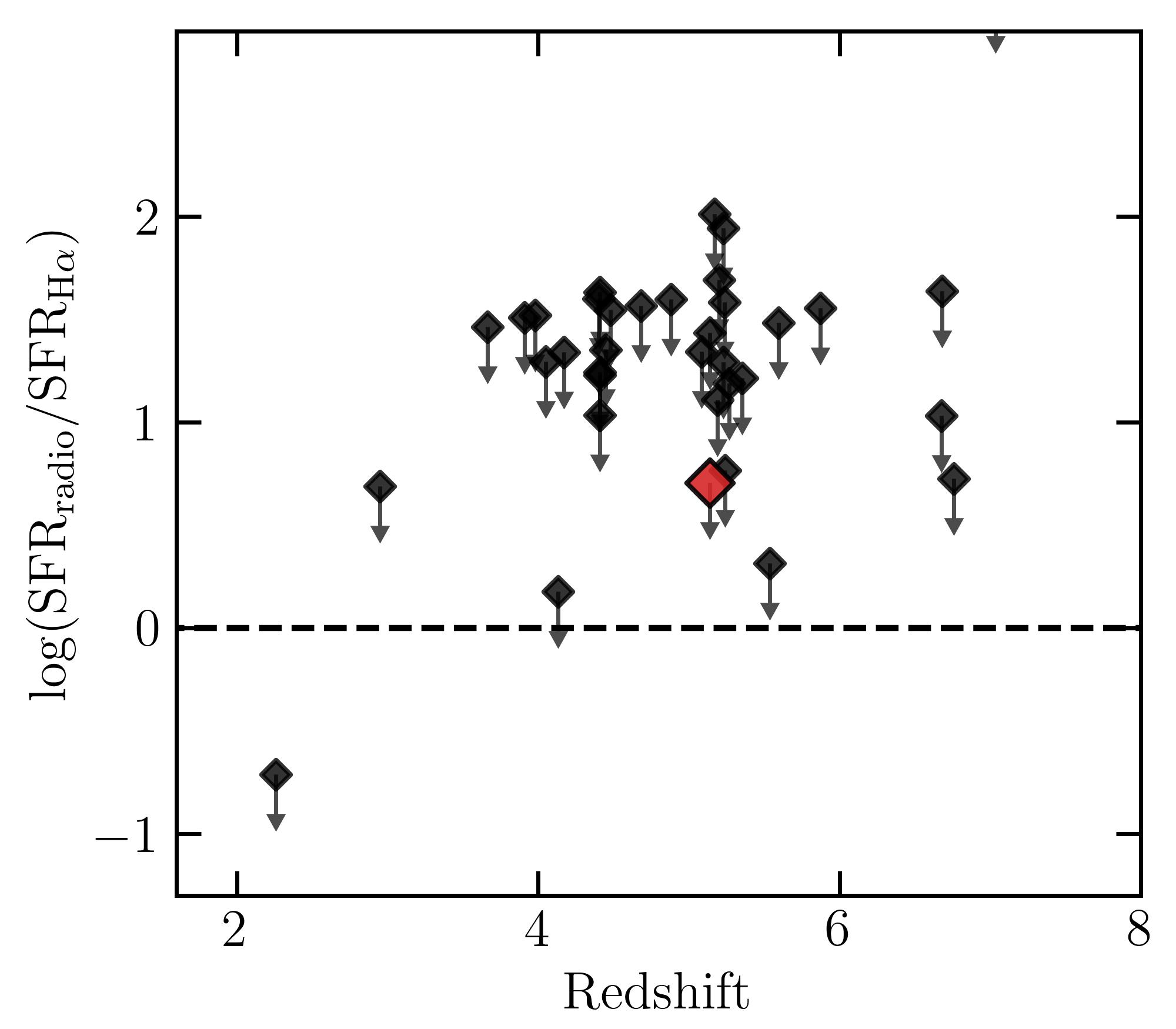}
    \caption{Ratio between the SFR derived from the radio luminosity upper limits and the SFR derived assuming the whole \Ha emission (narrow+broad) to be due to SF. The red data point corresponds to the value of the stack. The black dashed line traces the threshold below which the $SFR_{H\alpha}$ becomes incompatible with the radio undetection.}
    \label{fig:SFR_Ha}
\end{figure}

Some works also investigated alternative scenarios to the AGN one to explain the origin of the broad emission component, invoking for example dense star formation and outflows \citep{Kokubo2024}. Therefore, we computed the expected radio emission in the extreme hypothesis that the whole \Ha luminosity (broad+narrow) of these sources is due to SF. In particular, we used the relation reported in \cite{Reddy22}, derived from BPASS population synthesis models with metallicity $Z = 0.001$ (i.e., $\sim 0.07$ the solar metallicity, which is close to what typically inferred for these galaxies), including the effects of stellar binaries, and assuming an upper-mass initial mass function (IMF) cutoff of 100$M_{\odot}$. In Fig.~\ref{fig:SFR_Ha} we show the ratio between the SFR derived from the radio upper limits and the one obtained from the total \Ha luminosities. For most of the sources, the SFR derived from \Ha is not constraining to completely rule out scenarios alternative to AGN to explain the nature of the broad \Ha emission. However, there is one source, GN-28074, for which the upper limits are incompatible with the \Ha emission being due to SF (because $SFR_{data}/SFR_{H\alpha}<1$), while for other two BLAGN we have $SFR_{data}/SFR_{H\alpha}<2$ (and so they are close to being incompatible with \Ha due to SF). Using the relation of \cite{Kennicutt12} to derive the SFR from the \Ha luminosity, we obtained $\rm SFR_{H\alpha}$ 0.4 dex larger than those determined using \cite{Reddy22}, moving points downwards in Fig.~\ref{fig:SFR_Ha}. This is a further indication towards an AGN origin of the \Ha emission. 

\subsection{Simply radio-quiet AGN?} \label{sec:RQonly}
Most of the JWST-detected BLAGN show radio upper limits that are compatible with the expected RQ AGN luminosity predicted by the relations. 
The position of the stack in the four panels of Fig.~\ref{fig:RQ_panel} is consistent with the uncertainties and scatter of the four relations, but is generally placed below the 1:1 relation, possibly suggesting that these sources could be weaker than what expected from the standard AGN relations.  
In each panel, there is a group of sources, in particular GN-28074, whose upper limits are well below their expected luminosities. Given that the physical properties and the spectroscopic selection of these anomalously radio-weak BLAGN do not differ from the other sources investigated in this work, apart from the lower redshift or the larger $L_{H\alpha}$, and therefore $L_{bol}$), we might expect all these sources to have similar radio properties.

As we discussed in Sect.~\ref{sec:sfr}, the observed radio emission from these sources is not necessarily determined only by the AGN but also by the contribution of SF. On average, the AGN-driven radio luminosity is dominant over the SF one, yet this makes the observed radio upper limits conservative, given that the expected $L_{5GHz}$ (x-axis in Fig.~\ref{fig:RQ_panel}) refers only to the AGN-related component, and therefore we should subtract the SF radio contribution to the observed $L_{5GHz}$ (y-axis).

Furthermore, the points in Fig.~\ref{fig:RQ_panel} would move to the bottom right if a flatter spectral index is assumed \citep[$\alpha=0$, as typically seen in RQ AGN cores][]{Chen23, Alhosani22}, as shown in Appendix~\ref{sec:alpha0}. This would result in tighter radio weakness constraints.

{To further check the anomalous behaviour of our sources, we looked for non-JWST-selected BLAGN on the GOODS-N field falling on the same radio images considered in this work. In particular, we look at the type 1 AGN selection performed by \cite{Coil15} among the spectra of the MOSFIRE Deep Evolution Field (MOSDEF) survey. They identified 5 BLAGN at $z\sim2.5$ on GOODS-N, 4 of them are X-ray detected, while the undetected one has $\log L_{2-10keV}<42.88$. Given the well-established $L_{2-10keV}$-to-bolometric luminosity relation \citep{duras20}, these sources have $\log L_{bol}\sim 44-45$, comparable to those of our sources. However, in contrast with our findings, all these sources are radio-detected in at least one of the GOODS-N radio maps explored in this work. Additionally, the X-ray and radio luminosities of these sources perfectly align with the expectations from the $\rm L_{2-10keV}-L_{radio}$ relation of \cite{damato22}, as presented in Appendix~\ref{app:local_BLAGN}.}

Additionally, the radio (but also the X-ray) luminosity upper limits of the BLAGN investigated in this work are, in principle, compatible with the range of luminosities sampled by local RQ Low-Luminosity AGN \citep[LLAGN;][]{Panessa13,Baldi21}. However, most of the latter do not show broad Balmer emission lines, with some works suggesting that the BLR might disappear in these objects \citep{Nicastro00,Laor03}. Furthermore, the \OIII or \Ha luminosities of the JWST selected BLAGN are orders of magnitude larger than the corresponding values for typical broad-line LLAGN: the average $L_{H\alpha}$ of our sources is $\sim 1-5\times 10^{42}$ erg s$^{-1}$, while the typical range for broad-line LLAGN is $\rm L_{H\alpha}\sim 10^{37-39}$ erg s$^{-1}$ \citep{Ho97c,Balmaverde14}.\\

Therefore, to test the hypothesis that the 37 JWST-selected BLAGN are drawn from a standard RQ AGN population 
and that the fact that we do not detect any of them can be simply ascribed to the scatter of the relations described in Sect.~\ref{sec:Lradexp}, we performed a statistical test. For each of the relations reported in Fig.~\ref{fig:RQ_panel}, we generated the expected radio luminosities of the sample of 37 BLAGN and repeated this for $10^4$ times. For each source and each of the $10^4$ runs we randomly extract the $\rm L_{5GHz,exp}$ from a Gaussian distribution centered on the value returned by the relation and with $\sigma$ equal to the scatter of the relation. For the optical to radio luminosity relations (where we assumed two distinct radio loudness parameters) we considered $\sigma=0.42$ following \cite{Balokovic12}. Then, for each relation and for each run, we checked how many times all the 37 BLAGN had $\rm L_{5GHz,exp}<L_{5GHz,obs}$, i.e., they were all undetected. For the $\rm L_{[OIII]}-L_{radio}$ relation, we considered only the 13 sources for which $\rm L_{[OIII]}$ is available. The probabilities of non-detecting all the BLAGN are reported in Table~\ref{tab:probability} and are found to be negligible in all cases, except for the optical relation assuming $\log R=0$, where the probability is $\sim 25\%$. We note that the probabilities reported in Table~\ref{tab:probability} for the $L_{X}-L_{rad}$ and  FP relations, were computed by summing in quadrature the intrinsic scatter of the relations and the scatter of the relation used to compute the expected X-ray luminosity \citep{Jin12}.

In conclusion, the statistical test reported above suggests that the radio non-detection of all these uniformly selected sources is difficult to explain only in terms of the scatter of the RQ AGN relations. Therefore, {even if the RQ AGN nature cannot be completely ruled out,} in the following sections, we investigate possible scenarios that could produce a "radio-weakness" in these peculiar sources.

\begin{table}

    \centering
    \caption{Probabilities returned by the statistical test described in Sect.~\ref{sec:RQonly}}
    \begin{tabular}{ c  c c }
    \hline
    \hline
    Relation & Scatter [dex] & Probability\\
    \hline 
      $\rm L_X-L_{rad}$ & 0.5 & ${4\times 10^{-5}}$\\
      
      FP & 0.39 & ${<10^{-5}}$ \\
      
      $\rm L_{opt}-L_{rad},\ \log R=1$& 0.42 & ${<10^{-5}}$ \\
      
      $\rm L_{opt}-L_{rad},\ \log R=0$ & 0.42 & ${0.22}$ \\
      
      $\rm L_{[OIII]}-L_{rad}$ & 0.8 & ${4\times10^{-4}}$\\
      \hline

    \end{tabular}
    \tablefoot{ Each value corresponds to the probability that none of the BLAGN are detected in the 3GHz image of the GOODS-N field considering radio luminosities drawn from a Gaussian distribution centered on the expected value returned by the four relations described in Sect.~\ref{sec:Lradexp} and with $\sigma$ equal to each relation's scatter. The probabilities for the $L_{X}-L_{rad}$ and FP relations were computed considering the additional scatter of the relation used to compute the expected X-ray luminosity \citep[$\sigma=0.27$][]{duras20}}
    \label{tab:probability}
\end{table}

\subsection{Free-free absorption}\label{sec:FFA}

Radio emission is generally largely unaffected by obscuration and can be observed even from the most obscured environments, contrary to the X-rays that are almost completely absorbed at CTK hydrogen column densities \citep{Mazzolari24a}. However, also radio emission, especially at low frequency, can be absorbed in particular conditions as a consequence of radio free-free absorption due to the presence of a large density of free electrons \citep{Odea98,Baskin21}. One possibility (discussed in \citealt{Maiolino24_X}) to justify the extreme X-ray weakness of these sources, is that the X-ray photons are absorbed by a distribution of clouds with CTK column densities and a very low dust content, such as the Broad Line Region (BLR) clouds. Yet, the additional requirement is that their covering factor should be fairly high (close to 4$\pi$) if they are invoked to account for the large fraction of X-ray weak AGN discovered by the JWST.
Can these BLR clouds also be so thick (and with enough free electrons) to also absorb the radio emission? 

The free-free opacity $\tau_{ff}$ is given by:
\begin{equation}\label{eq:tau_ff_ne}
    \rm \tau_{ff}=3.28\  10^{-7}\times \biggl(\frac{T_e}{10^4 K}\biggr)^{-1.35}
    \biggl(\frac{\nu_{rest}}{GHz}\biggr)^{-2.1}
    \biggl(\frac{EM}{pc\ cm^{-6}}\biggr)
\end{equation}
where $T_e$ is the medium electron temperature, $\nu_{rest}$ is the rest frame frequency of the radio emission (i.e. $\nu_{rest}=(1+z)\nu_{obs}$), and $EM$ is the emission measure, which is defined as the integral of the square of the free electrons density $n_e^2$ along the line of sight. The EM can be easily expressed
in terms of the  
$f_i^2N_H^2$
column of free electrons (or, equivalently, the column of ionized hydrogen), in the form of $N_e^2~r^{-1}$
where $r$ the thickness of the ionized medium along the line of sight. Therefore, the expression of $\tau_{ff}$ becomes:
\begin{equation}\label{eq:tau_ff_NH}
   \rm  \tau_{ff}=3 \times   \ \biggl(\frac{N_e}{10^{22} cm^{-2}}\biggr)^2
    \biggl(\frac{T}{10^4 K}\biggr)^{-1.35}
    \biggl(\frac{\nu_{rest}}{GHz}\biggr)^{-2.1}
    \biggl(\frac{r}{1 pc}\biggr)^{-1},
\end{equation}
in the source rest frame. 
The BLR clouds usually have a temperature in the range $5000-25000$K \citep{Petersonbook}. 
Following the computation in \cite{Netzer13}, typical column densities of ionized gas in AGN BLR are of the order of $\rm 1-5\times 10^{22} cm^{-2}$. As another example, \cite{Ferland09} found for the local AGN Akn120 an ionized column density of $\rm \sim 5\times 10^{21} cm^{-2}$.
The estimated thicknesses of the ionized component of the BLR clouds is generally very small. \citet{Petersonbook}, simply requiring photoionization equilibrium of the clouds, derived $r\sim 10^{-6}$ that can become $r\sim 10^{-3}$ if the whole BLR is ionized. Therefore, even conservatively assuming $N_e\sim 5\times 10^{21} cm^{-2}$, we get $\tau _{ff}>300-3\times 10^5$ at $\nu_{obs}=$1.5\, GHz, implying large obscuration. We note that the possibility of radio free-free absorption had already been pointed out for the `Rosetta Stone' (GN-28074; \citealt{Juodzbalis24_rosetta}). 

The effect of free-free absorption on the radio emission spectrum is to progressively flatten the low-frequency part and eventually even produce an inverted radio spectrum on that side. Interestingly, this radio spectral shape, with a significant flattening at $\nu< 1.4$\,GHz was observed in SBS 0355-052 \citep{Johnson09}; the local radio weak dwarf AGN reported in Fig.~\ref{fig:RQ_panel}.

However, free-free absorption would imply that the whole (or at least most of the) radio emission is confined within scales that are smaller or comparable to those of the absorbing medium. If the absorbing medium is made of BLR gas clouds, then we have to check whether the radio emission is produced on larger or smaller scales than those of the BLR. To estimate the radi of the BLR for our sources we used the scaling relation between the BLR radius and AGN luminosity derived in \cite{GRAVITY_BLR} using Near-infrared interferometry. Based on this relation, the median BLR size in our sample is $\sim 0.03$pc and ranges $0.01<R_{BLR}/pc<0.07$. We then derived the minimum size of the radio-emitting region, following \cite{Laor08}. We can make the reasonable hypothesis of an inner optically thick radio-emitting source and compute this size considering the maximum flux per unit area emitted by a synchrotron source. Indeed, the presence of an inner, compact, and with flat spectral slope radio core was observed in many RQ AGN using high-resolution interferometry \citep{Chen23,Paul24,Laor08,Alhosani22}. With the appropriate corrections for high-redshift sources, the relation derived in \cite{Laor08} has the form:
\begin{equation}\label{eq:Rrad}
    \rm \frac{R_{\rm rad}}{pc}=0.47\ \biggl(\frac{L_{\nu,rest}}{10^{30} erg\ s^{-1} Hz^{-1}}\biggr) ^{0.4} \
    \biggl(\frac{L_{bol}}{10^{46} erg s^{-1} }\biggr) ^{0.1}
    \biggl(\frac{\nu_{rest}}{GHz}\biggr) ^{-1}.
\end{equation}
In the equation above, $L_{\nu,rest}$ is the radio luminosity at rest-frame frequency $\nu_{rest}$, and $L_{bol}$ is the bolometric luminosity. Given that we are testing the radio absorption scenario, for each of the sources we assume as $L_{\nu,rest}$ the median radio AGN luminosity obtained from the relations described in Sect.~\ref{sec:Lradexp}. 
Considering the equation above, the minimum size of the radio-emitting regions for our sources is on average comparable or lower to that of the BLR, having on average an extension $R<0.01$pc for $\nu_{obs}>1.5\,GHz$.

Even if with the discussion above we found that the minimum scale of the radio emission can reside inside the BLR, some recent works show that the bulk of the radio emission, also in RQ AGN, can extend up to scales much larger than $\sim 1 pc$. \cite{Chen23}, comparing high-resolution and low-resolution radio observations of a sample of 71 local RQ AGN (corresponding to physical scales of 2-20 pc and 200-2000 pc, respectively), found that the average radio luminosity associated with the compact core is one order of magnitude lower than the one associated to the extended scale radio emission, suggesting that only $\sim 10\%$ of the radio emission of these sources in confined in their innermost regions. They also found that for most objects, the compact cores show flatter spectral indices ($\alpha>-0.5$) compared to the extended radio emission.

Therefore, one might wonder if free-free absorption can also happen at larger scales, for example up to the torus scale. We have now to apply Eq.~\ref{eq:tau_ff_NH} to the typical AGN torus conditions. Typical torus temperatures are of the order of $T_e\sim 10^3$K, and it can be considered made by two distinct components, a molecular zone and an atomic zone. In the first, the amount of free electrons is completely negligible \citep{Risaliti03}. For the atomic zone, the lower temperatures and pressures of the torus compared to the BLR (together with the fact that most of the ionizing radiation might be absorbed at the BLR scale) determine the fraction of ionized hydrogen to be lower, therefore suppressing $\tau_{ff}$ in  Eq. \ref{eq:tau_ff_NH}.
Therefore, if free-free absorption is at the origin of a radio weakness in our sources, it likely happen on scales that are comparable to those of the BLR, or its vicinity, providing an upper limit also to the extension of the radio emission. It is worth noting that for GN-28074, based on photoionization calculations, \cite{Juodzbalis24_rosetta} estimate a location of the absorbing medium that is outside the typical BLR radius (though still within the inner side of the torus as defined by the dust sublimation radius), suggesting that the absorber is actually dense gas located in the outer regions of the BLR, although still confined within a very compact zone.\\

Another possible scenario to explain the possible radio weakness of our sources is Synchrotron self-absorption (SSA). However, it is hard to test with the data in our hand and is only briefly discussed in Appendix~\ref{sec:app3}.

\subsection{Lack (or disruption) of magnetic field and X-ray corona}\label{sec:magnetic_field}
A magnetic field is essential to observe AGN radio emission since synchrotron emission relies on electrons accelerated by this field and interacting with it. The X-ray corona, producing most of the SMBH’s X-ray emission, is also linked to magnetic processes \citep{Jafari19,merloni03,Merloni02}. Indeed, the corona is thought to be a magnetically-powered plasma, with magnetic energy heating and sustaining electrons near the SMBH. Therefore, the presence of the nuclear magnetic field, or lack thereof, can link the strength of X-ray and radio emission.

The origin of the magnetic field close to an SMBH is generally ascribed to magneto-rotational instabilities in the SMBH accretion disk, which produce a seed magnetic field that is then amplified and transported vertically due to different possible mechanisms, such as buoyancy \citep{Merloni02,Miller00}. Following the description provided in \cite{Dimatteo98}, magnetic field amplification as a consequence of buoyancy can lead to the formation of an accretion-disk corona consisting of many magnetic loops where the magnetic field intensity is set by the condition:
\begin{equation}\label{eq:Bcorona}
    \rm \frac{B_{corona}^2}{8\pi}=n_{disk}KT_{disk},
\end{equation}
where $n_{disk}$ is the gas number density in the disk and $T_{disk}$ the disk temperature. Then, magnetic reconnection episodes can realese significant amount of energy and heat this region up to usual X-ray emitting temperatures. Taking the equation for the maximum disk temperature reported in \cite{Bonning07} for a Schwarzschild SMBH, and assuming $n_{disk}\sim 10^{16}\ \rm cm^{-3}$, we found an average value of $B_{corona}\sim 2\times 10^3$~G for the BLAGN investigated in this work. The estimated coronal magnetic field is consistent within a factor of a few with the equipartition magnetic field derived in the assumption of optically thick radio emission:
\begin{equation}\label{eq:Beq}
    \rm \frac{B_{eq}^2}{8\pi}=\frac{L_{bol}}{4\pi R^2 c}\ G,
\end{equation} 
when rescaled to the size of the X-ray corona ($R\sim$10$R_g$). This may support the link between the coronal magnetic field and the one giving origin to the radio emission, if standard conditions are present.

In principle, if the magnetic field strength or its pattern are significantly altered, the magnetic heating of the corona and the production of synchrotron emission in the innermost AGN region can both cease, determining a simultaneous suppression of the radio and the X-ray emissions. Similarly, in the hypothesis of a coronal-related origin of the synchrotron emission \citep{Laor08,panessa19}, the lack (or a significant cooling) of the electrons constituting the X-ray corona can determine a similar fate. 

With the data in our hands it is not possible to verify this scenario, and a proper treatment of the magnetic properties of these objects is well beyond the scope of this paper. In the following, we limit the discussion to the comparison of the properties of our sources with scenarios that have been recently discussed in the literature.

One scenario that can induce the above mentioned effects can be a much lower inner disk or coronal plasma temperature. This can happen, for example, as a consequence of a phase of super-Eddington accretion that can empty the innermost (and hotter) region of the accretion disk and generate a cooling of the electron plasma, possibly leading to the weakening of the magnetic field or of the X-ray corona \citep{Lupi24, Madau2024_SE}.
The mildly super-Eddington scenario was proposed by \cite{Pacucci24} to explain the widespread population of X-ray weak AGN identified by the JWST. The authors used general relativistic radiation magnetohydrodynamics (GRRMHD) simulations to investigate the origin of the JWST selected BLAGN X-ray weakness, concluding that episodes of super-Eddington accretion can cause the inner part of the accretion disk to thicken, effectively shielding the most energetic photons, thus leading to a reduced X-ray output. They also observed that X-ray weak sources differ from typical X-ray AGN primarily in the temperature of the inner accretion disk and in disk jet production, with X-ray weak sources exhibiting lower disk temperatures and producing less energetic jets. \cite{Madau2024_SE}, investigating the same topic assumed a two-phase disk/corona scenario and super-Eddington accretion, finding that the X-ray corona can be embedded in a funneled geometry, where soft photons can produce a significant cooling of the coronal plasma down to much lower temperatures than for standard accretion rates, i.e. $kT_e\lesssim 30 eV$. This consequently produces an extremely soft X-ray spectrum and, therefore, an intrinsic X-ray weakness. This cooling of the coronal electron plasma can significantly impact also the efficiency of synchrotron emission at comparable scales, possibly affecting the radio emission also at the larger scales probed by our frequencies (the minimum radius of the radio emitting region is at least $100$ times larger than the X-ray corona). 

A super-Edditgnton origin of the X-ray weakness was also suggested by \cite{Paul24} for the X-ray weak sources shown in Fig.~\ref{fig:RQ_panel}, and some of them are indeed also radio-weak. NLS1 are generally characterized by high-accretion rates and high X-ray to bolometric corrections \citep{Vasudevan07}, and some of these are indeed observed to be radio-weak (see Sect.~\ref{sec:radio_weakness}). However, the NLS1 scenario alone appears not sufficient to justify some of the X-ray weakest JWST-detected BLAGN \citep{Maiolino24_X} and it is not possible to attribute to NLS1 a uniform radio emission behavior as shown by \cite{Berton16,Berton18}.

In \cite{Arcodia24}, the authors investigated the possible lack of an X-ray corona in a sample of UV-optical-IR variability selected SMBH in dwarf galaxies observed by eROSITA. These sources show a significant observed X-ray weakness compared to their optical luminosity, and most of them are also radio-weak with respect to the \cite{merloni03} FP relation, further suggesting that SMBH in dwarf galaxies can be much less efficient in sustaining a magnetically powered corona.

\section{Conclusions}
We have analyzed the radio emission from a sample of 37 JWST-selected Broad Line AGN residing in the GOODS-N field considering a wide set of radio images at multiple frequencies (144\,MHz, 1.5\,GHz, 3\,GHz, 5.5\,GHz, 10\,GHz).
Our findings are summarized in the following.
\begin{itemize}
    \item None of the 37 BLAGN has a counterpart in any of the radio images covering the GOODS-N field. Even the radio stacking analysis did not show any indication of radio emission. We then used, as a reference, the radio upper limits for the single sources and for the stack obtained from the 3\,GHz image, which provides the deepest constraints. In particular, the $3\sigma$ upper limit from the stack corresponds to a rest frame 5\,GHz radio luminosity $\rm \log L_{5GHz} \; (erg \; s^{-1})=39.4$.
    
    \item  The upper limits in the observed radio luminosity are compared with the expected radio luminosities for RQ AGN derived by using a set of different relations, including $L_{[OIII]}-L_{radio}$, $L_{X}-L_{radio}$, $L_{optical}-L_{radio}$, and FP relations. {This comparison shows that 
    most of the individual BLAGN, as well as their stack,  are compatible with being RQ AGN, considering the scatter of the relations. However, this comparison also showed that some of these sources significantly deviate from the expected 1:1 relation, potentially suggesting a weaker radio emission. } 
    
    \item We investigate the SF contribution to the radio emission of these objects, finding that, for most of them, the SF-driven radio emission is generally sub-dominant compared to the expected RQ AGN radio emission but yet not negligible for some sources. This makes these BLAGN observed radio luminosity upper limits even more stringent, given that we should take into account only the AGN-related radio emission component in Fig.~\ref{fig:RQ_panel}. Moreover, we computed the SFR assuming the case in which the whole \Ha emission (broad + narrow) is caused by SF, and we found that for a few sources, this is incompatible with the radio non-detection. This further supports the AGN origin (BLR) of the broad H$\alpha$ in these sources.
    
    \item We performed a statistical test to check the hypothesis that not detecting any of the 37  investigated sources is still statistically consistent with the RQ AGN scenario and can be simply ascribed to the scatter of the relations considered. The test revealed probabilities $\rm P\sim 10^{-4}$ for all the relations except for one, suggesting that these BLAGN may instead be characterized by a radio-weakness (in addition to the observed X-ray weakness). {Additionally, non-JWST selected BLAGN on the GOODS-N field \citep{Coil15} are clearly detected in the radio images considered in this work, contrary to our objects.} Therefore, we discussed possible scenarios that can produce a radio weakness.
    
    \item A Compton-thick and almost spherical distribution of the BLR clouds has been proposed as a possible cause of the observed X-ray weakness of these sources. We investigate the possibility that this distribution can also entail free-free absorption of the radio emission. We find that free-free absorption is possible provided that most of the radio emission is confined inside the scale of the BLR or its vicinity (i.e., on scales smaller than the sublimation radius).

    \item We also argue that the lack or disruption of the inner magnetic field and hot corona, possibly associated with Super Eddington accretion, can simultaneously explain an intrinsic radio weakness and an intrinsic X-ray weakness.
    
\end{itemize}

The analysis performed in this work shows that to definitively probe the nature, the origin, and possibly the weakness/lack of the radio emission of these sources, deeper radio images are needed. In particular, having radio images one order of magnitude deeper than those provided by the current facilities will allow us to set more stringent constraints on the radio properties of JWST-selected X-ray weak AGN. Sub-arcsec resolution will also be critical to ensure that contamination from galaxy-scale radio emission is low. In this sense, the deep radio continuum observations that will be performed by the SKAO will play an important role, given that they are expected to reach an rms of $0.05 \; (0.04) \mu$Jy at 0.5 (0.1) arcsec resolution in the UD surveys at frequencies $\sim 1$ (10) GHz \citep{prandoni15}.

   

\begin{acknowledgements}
GM acknowledges useful conversations and discussions with A. Laor, C. Spingola, G. Cresci, A. Trinca. We acknowledge support 
from the Bando Ricerca Fondamentale INAF 2022 and 2023. ID acknowledges funding by the European Union – NextGenerationEU, RRF M4C2 1.1, Project 2022JZJBHM: "AGN-sCAN: zooming-in on the AGN-galaxy connection since the cosmic noon" - CUP C53D23001120006. E.F.-J.A. acknowledge support from UNAM-PAPIIT project IA102023, and from CONAHCyT Ciencia de Frontera project ID: CF-2023-I-506. The National Radio Astronomy Observatory is a facility of the National Science Foundation operated under
cooperative agreement by Associated Universities, Inc.
RM, XJ, FD, JS and IJ acknowledge support by the Science and Technology Facilities Council (STFC), by the ERC through Advanced Grant 695671 ”QUENCH”, and by the UKRI Frontier Research grant RISEandFALL. RM also acknowledges funding from a research professorship from the Royal Society.
MS acknowledges financial support from the Italian Ministry for University and Research, through the grant PNRR-M4C2-I1.1-PRIN 2022-PE9-SEAWIND: Super-Eddington Accretion: Wind, INflow and Disk-F53D23001250006-NextGenerationEU.
IP acknowledges support from INAF under the Large Grant 2022 funding scheme (project "MeerKAT and LOFAR Team up: a Unique Radio Window on Galaxy/AGN co-Evolution")".
H\"U acknowledges support through the ERC Starting Grant 101164796 ``APEX''.

\end{acknowledgements}

\bibliographystyle{aa}
\bibliography{literature}

\begin{appendix} 

\section{Discarding RL AGN scenario}\label{sec:RL}
\begin{figure}[h!]
    \centering
    \includegraphics[width=1\linewidth]{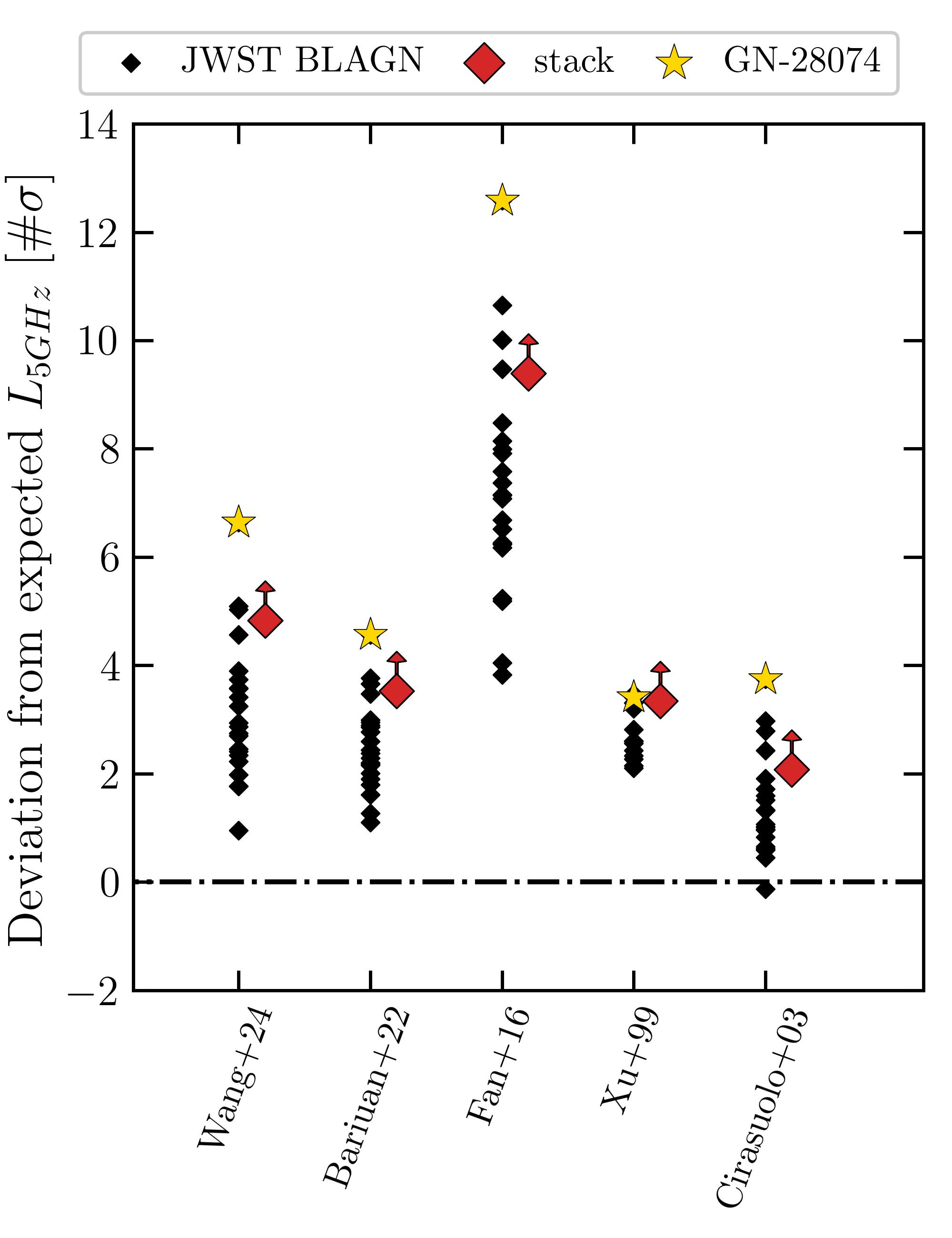}
    \caption{Deviation of the observed radio luminosity upper limits of the 37 JWST-detected BLAGN (black points) and of their stack (red diamond) compared to their expected radio luminosity according to the different RL AGN luminosity relations reported on the x-axis. The deviation is expressed in numbers of $\sigma$, where $\sigma$ is the intrinsic scatter of each relation.\cite{Wang24_FP} and \cite{Bariuan22} are FP relations, \cite{Fan16} is a direct X-ray to radio luminosity relation, \cite{Xu99} is a $\rm L_{[OIII]} - L_{rad}$  relation, while \cite{Cirasuolo03} is a $\rm L_{opt}-L_{rad}$ relation. We did not put the arrow indicating the upper limits of each of the BLAGN for clarity.  }
    \label{fig:RL}
\end{figure}

In this section, we show that the radio undetection of the sources investigated in this work is largely inconsistent with their expected radio luminosities if they are assumed to be RL. In Fig.~\ref{fig:RL}, we show the deviation, expressed in number of $\sigma$ (where $\sigma$ is the intrinsic scatter of the relations), of the radio upper limits derived from the 3GHz radio image from the expected RL AGN radio luminosity. For every RL AGN luminosity relation the deviation of the stack is $\gtrsim 2.5 \sigma$.\\

\section{Observed radio luminosity upper limits versus observable quantities}\label{app:RQ_obs}
{Here we show the analog of Fig.~\ref{fig:RQ_panel} but directly comparing the observed radio luminosity upper limits to the available observable quantities used to compute the expected radio luminosities in Sect.~\ref{sec:Lradexp}. Data points are compared directly with the scaling relations presented in Sect.~\ref{sec:Lradexp}.}
\begin{figure}[h!]
    \centering
    \includegraphics[width=1\linewidth]{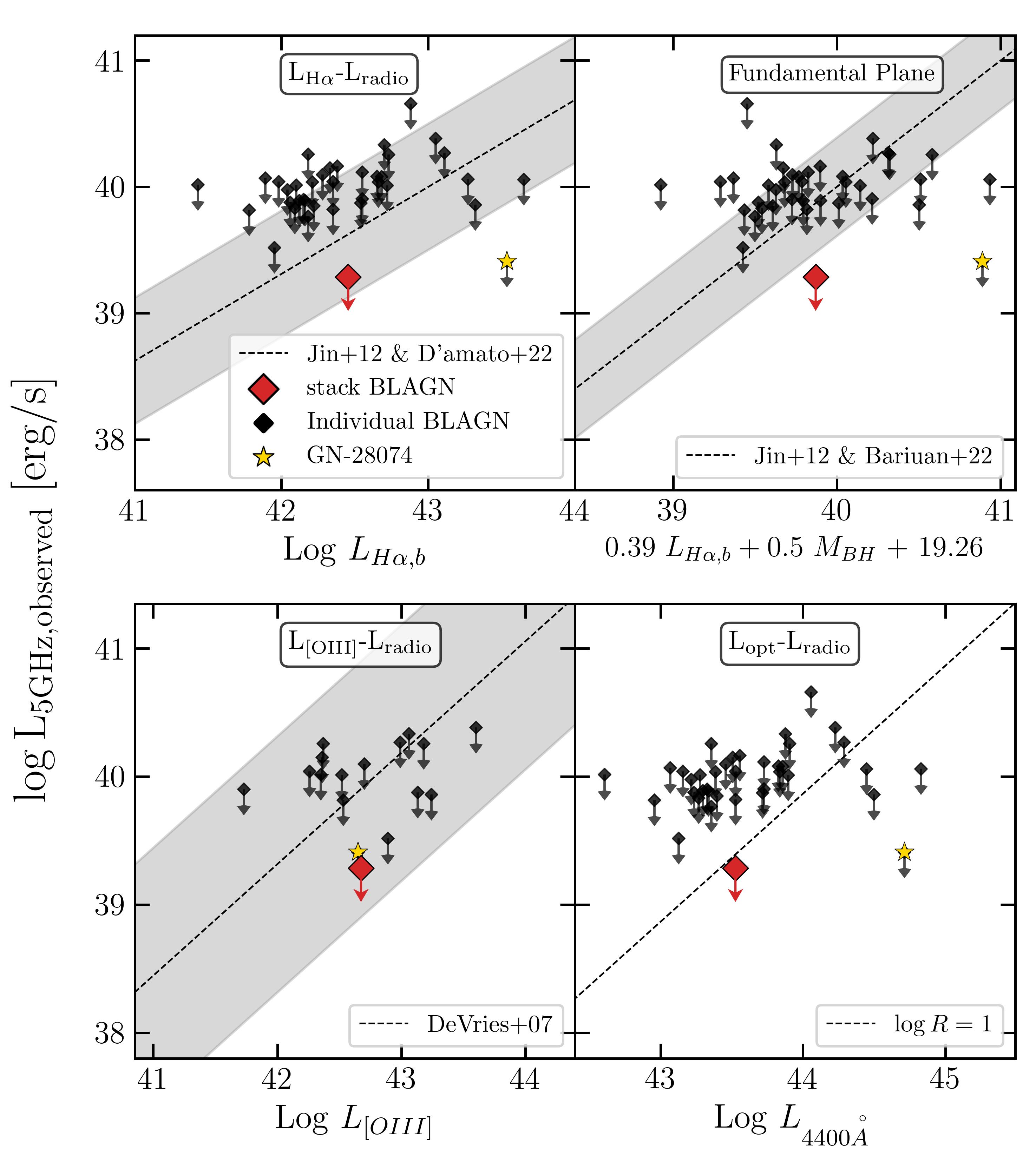}
    \caption{Similar to Fig.\ref{fig:RQ_panel}, but directly showing the observed rest-frame 5GHz radio luminosity versus the observable quantities used to compute the expected radio luminosities in Sect.~\ref{sec:Lradexp}. In particular, the observed radio luminosities are compared with the broad \Ha luminosities (upper left), the linear combination of broad \Ha luminosity and $M_{BH}$ derived from the fundamental plane of \cite{Bariuan22} (top right), the \OIIIl5007 luminosities (bottom left) and the 4400Å luminosities (bottom right). Data points for single BLAGN (black) and for their stack (red) are compared to the scaling relations described in Sect.~\ref{sec:Lradexp}. In the two top panels we combined the $L_{H\alpha, broad}$ vs $L_{\rm 2-10 keV}$ relation from \cite{Jin12} with the $L_{\rm 2-10 keV}$ vs $L_{radio}$ relation of \cite{damato22} (top left), and the fundamental plane relation of \cite{Bariuan22} (top rigth).}
    \label{fig:RQ_panel_referee}
\end{figure}

\section{Compact radio sources with $\alpha=0$}\label{sec:alpha0}
\begin{figure}[h!]
    \centering
    \includegraphics[width=1\linewidth]{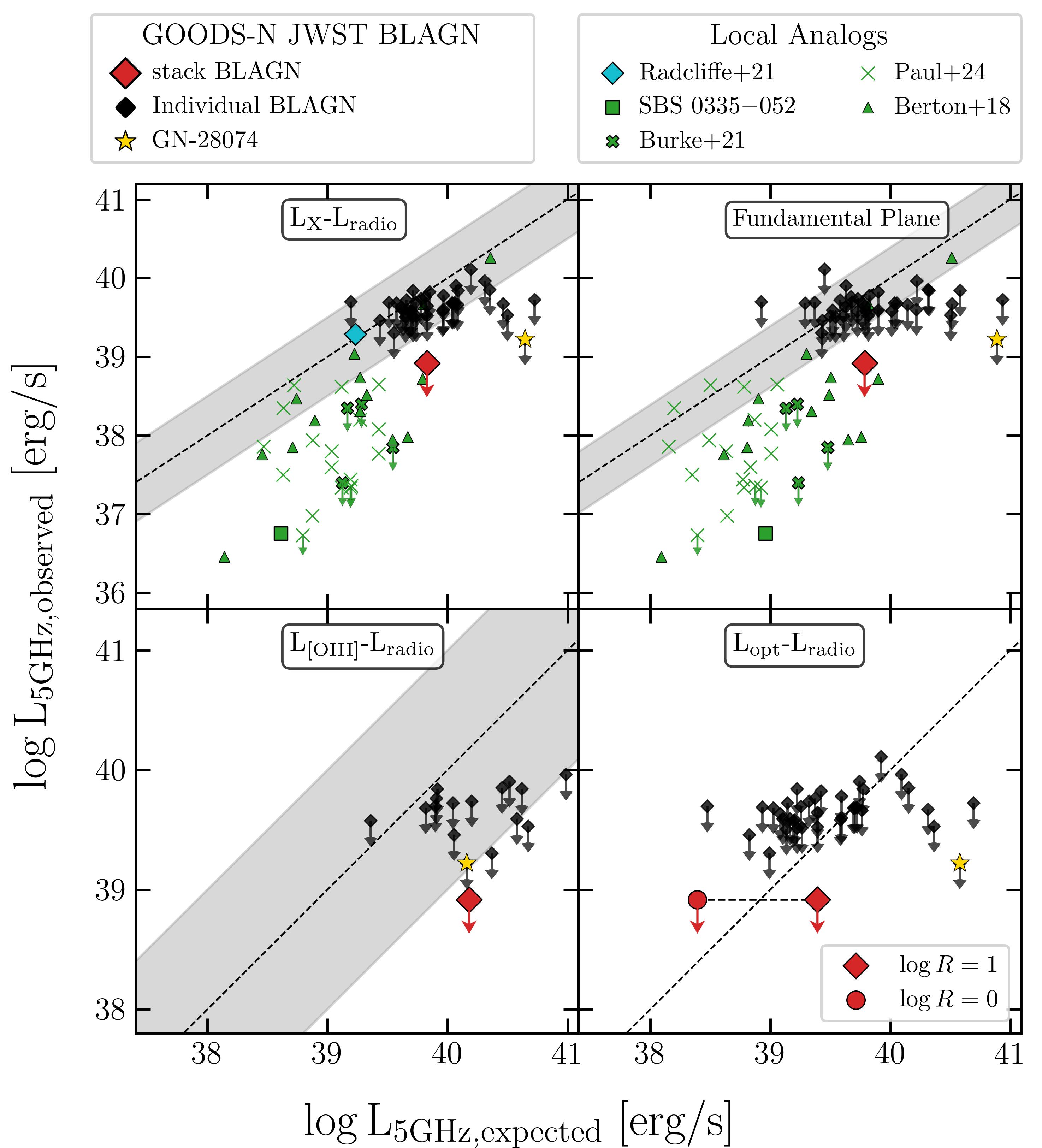}
    \caption{Same as in Fig.\ref{fig:RQ_panel}, but considering $\alpha=0$ and taking the 10GHz radio image, that for $\alpha=0$ provide deeper constrains at 5GHz than the 3GHz image. }
    \label{fig:RQ_panel_a0}
\end{figure}

If we assume a central compact radio-emitting source with a flat radio spectral slope, $\alpha=0$ (instead of $\alpha=-0.5$ assumed in the main text), as observed for the cores of many RQ AGN \citep{Chen23,Alhosani22}, Fig.~\ref{fig:RQ_panel} translates into Fig.~\ref{fig:RQ_panel_a0}, where the radio upper limits are more stringent and the deviation from the expected radio luminosity is much more significant.

\section{Non-JWST-selected BLAGN on GOODS-N}\label{app:local_BLAGN}
{As discussed in Sect.~\ref{sec:RQonly}, we investigated the radio properties of $z\sim 2-3$ BLAGN on the GOODS-N field identified among the MOSFIRE spectra. Here we show that, contrary to the JWST-selected BLAGN, these sources have radio luminosities perfectly consistent with the RQ AGN relation from \cite{damato22}, if not even in excess.}
\begin{figure}[h!]
    \centering
    \includegraphics[width=1\linewidth]{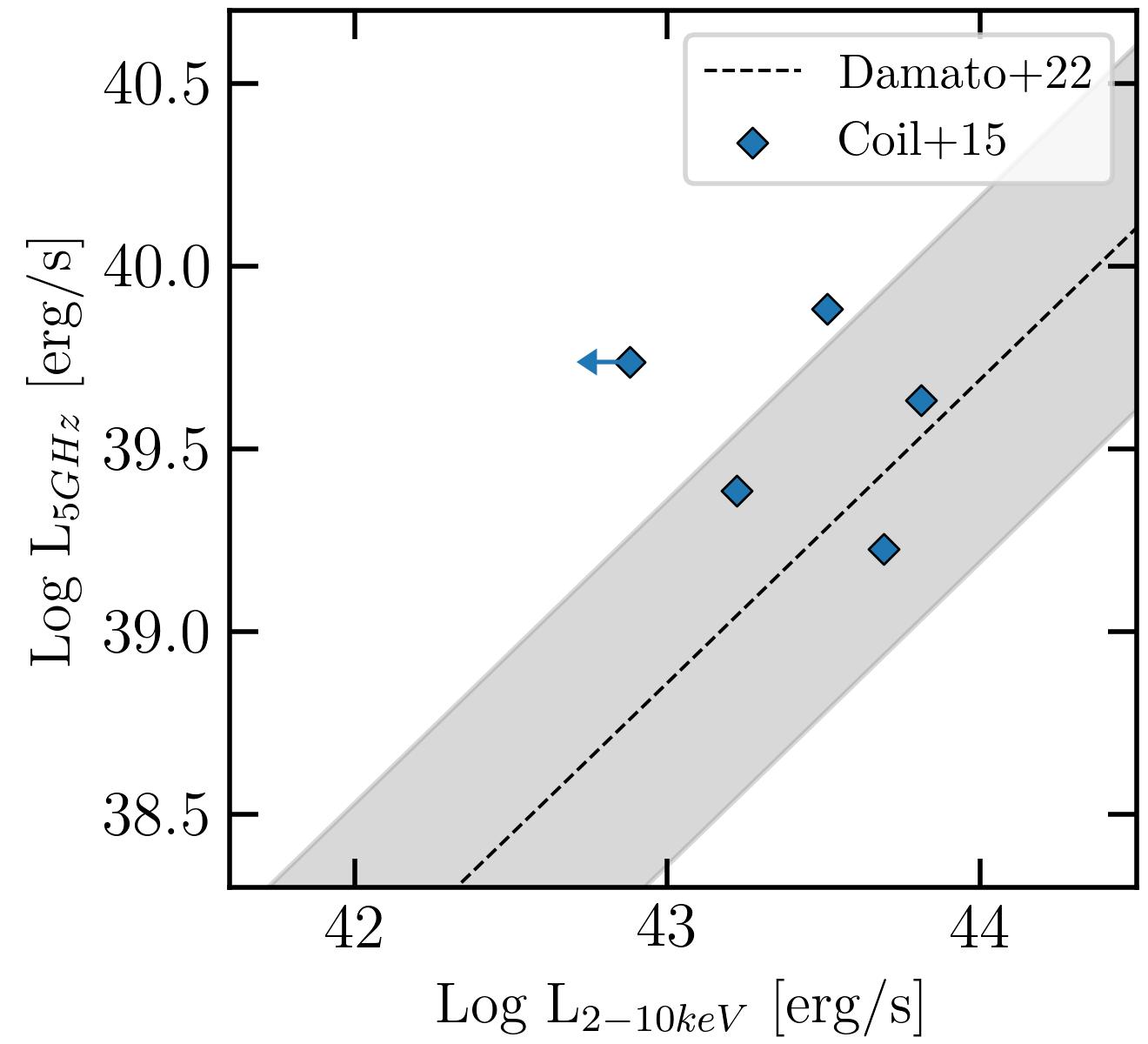}
    \caption{Radio versus X-ray luminosity distribution of the 5 BLAGN presented in \cite{Coil15}. The observed radio luminosities are consistent with the prediction of the \cite{damato22} relation except for one (the X-ray undetected one) being probably radio loud.
}
    \label{fig:enter-label}
\end{figure}

\section{Synchrotron self absorption}\label{sec:app3}
Similarly to FFA, also synchrotron self-absorption (SSA) can lead to a substantial decrease in the expected radio emission, even if the physical processes are different. Indeed, when synchrothron emission is self absorbed, the observed radio spectrum shows a turnover and a steepening below the so-called turnover (or peak) frequency $\nu_{SSA}$, which depends on the strength of the magnetic field.

Given the undetection of the sources (and therefore the lack of information on the peak frequency, on the peak flux, and on the extension of the radio emission), we do not have the possibility to reliably test the SSA hypothesis scenario.
However, using the \cite{Orienti13} relation between SSA peak frequency and the largest linear size of the radio emission observed for a sample of compact and young radio sources,  we can tentatively derive the maximum size of the radio emitting region, under the hypothesis that the peak frequency is below the frequency range explored by us (i.e. above  $\nu_{obs}=10\, {\rm GHz}$).
We derived for our sources maximum sizes in the range $1-0.1$pc, i.e. consistently larger than the minimum radio-emitting region, obtained assuming the same frequency from Eq.~\ref{eq:Rrad}, but still quite  confined.

\end{appendix}

\end{document}